\documentclass[sn-mathphys-num]{sn-jnl}% Math and Physical Sciences Numbered Reference Style 
\usepackage{siunitx}
\usepackage{graphicx}%
\usepackage{caption}
\usepackage{multirow}%
\usepackage{amsmath,amssymb,amsfonts}%
\usepackage{amsthm}%
\usepackage{mathrsfs}%
\usepackage[title]{appendix}%
\usepackage{xcolor}%
\usepackage{textcomp}%
\usepackage{manyfoot}%
\usepackage{booktabs}%
\usepackage{algorithm}%
\usepackage{algorithmicx}%
\usepackage{algpseudocode}%
\usepackage{listings}%
\usepackage{paralist}
\usepackage{array}
\usepackage[export]{adjustbox}
\usepackage{ulem}
\usepackage{csquotes}
\usepackage{lineno}
\usepackage{makecell}
\theoremstyle{thmstyleone}%
\theoremstyle{thmstyletwo}%

\theoremstyle{thmstylethree}%

\newcommand{\etal}{\textit{et al.}}
\newcommand{\ie}{\textit{i.e.}}
\newcommand{\eg}{\textit{e.g.}}
\makeatletter
\DeclareRobustCommand\onedot{\futurelet\@let@token\@onedot}
\def\@onedot{\ifx\@let@token.\else.\null\fi\xspace}
\begin{document}

\title[Article Title]{\textcolor{black}{Virtual Fluoroscopy for Interventional Guidance using Magnetic Tracking}}

%%=============================================================%%
%% GivenName	-> \fnm{Joergen W.}
%% Particle	-> \spfx{van der} -> surname prefix
%% FamilyName	-> \sur{Ploeg}
%% Suffix	-> \sfx{IV}
%% \author*[1,2]{\fnm{Joergen W.} \spfx{van der} \sur{Ploeg} 
%%  \sfx{IV}}\email{iauthor@gmail.com}
%%=============================================================%%

\author*[1,2]{\fnm{Shuwei} \sur{Xing}}\email{xshuwei@uwo.ca}

\author[1,4]{\fnm{Inaara} \sur{Ahmed-Fazal}}
\email{iahmedfa@uwo.ca}

%\email{iahmedfa@uwo.ca}
\author[5]{\fnm{Utsav} \sur{Pardasani}}
\email{upardasani@ndigital.com}
%\equalcont{These authors contributed equally to this work.}
\author[5]{\fnm{Uditha} \sur{Jayarathne}}\email{ujayarathne@ndigital.com}
\author[5]{\fnm{Scott} \sur{Illsley}}\email{sillsley@ndigital.com}
% \author[1]{\fnm{XX} \sur{XX}}
%\email{XXX}
%\equalcont{These authors contributed equally to this work.}
\author[1,2,3]{\fnm{Aaron} \sur{Fenster}}
\email{afenster@uwo.ca}

\author[1,2,3]{\fnm{Terry M.} \sur{Peters}}
\email{tpeters2@uwo.ca}

\author[1,2,3,4]{\fnm{Elvis C.S.} \sur{Chen}}
\email{chene@robarts.ca}

\affil[1]{\orgdiv{Robarts Research Institute}, \orgname{Western University}, \orgaddress{\street{100 Perth St.}, \city{London}, \postcode{N6A 5B7}, \state{ON}, \country{Canada}}}

\affil[2]{\orgdiv{School of Biomedical Engineering}, \orgname{Western University}, \orgaddress{\street{100 Perth St.}, \city{London}, \postcode{N6A 5B7}, \state{ON}, \country{Canada}}}

\affil[3]{\orgdiv{Department of Medical Biophysics}, \orgname{Western University}, \orgaddress{\street{100 Perth St.}, \city{London}, \postcode{N6A 5B7}, \state{ON}, \country{Canada}}}

%\affil[4]{\orgdiv{Department of Medical Imaging}, \orgname{Western University}, \orgaddress{\street{100 Perth St.}, \city{London}, \postcode{N6A 5B7}, \state{ON}, \country{Canada}}}

\affil[4]{\orgdiv{Department of Electrical and Computer Engineering}, \orgname{Western University}, \orgaddress{\street{100 Perth St.}, \city{London}, \postcode{N6A 5B7}, \state{ON}, \country{Canada}}}

%\affil[6]{\orgname{Northern Digital Inc.}, \orgaddress{\street{103 Randall Dr.}, \city{Waterloo}, \postcode{N2V 1C5}, \state{ON}, \country{Canada}}}

\affil[5]{\orgname{Northern Digital Inc.}, \orgaddress{\street{103 Randall Dr.}, \city{Waterloo}, \postcode{N2V 1C5}, \state{ON}, \country{Canada}}}

%%==================================%%
%% Sample for unstructured abstract %%
%%==================================%%

\abstract{\textbf{Purpose:} \textcolor{black}{In conventional fluoroscopy-guided interventions, the 2D projective nature of X-ray imaging limits depth perception and leads to prolonged radiation exposure. Virtual fluoroscopy, combined with spatially tracked surgical instruments, is a promising strategy to mitigate these limitations. While magnetic tracking shows unique advantages, particularly in tracking flexible instruments, it remains under-explored due to interference from ferromagnetic materials in the C-arm room. This work proposes a virtual fluoroscopy workflow by effectively integrating magnetic tracking, and demonstrates its clinical efficacy.}
\textbf{Methods:} An automatic virtual fluoroscopy workflow was developed using a radiolucent tabletop field generator prototype. Specifically, we developed a fluoro-CT registration approach with automatic 2D-3D shared landmark correspondence to establish the C-arm-patient relationship, along with a general C-arm modelling approach to calculate desired poses and generate corresponding virtual fluoroscopic images.
\textbf{Results:} Testing on a dataset with views ranging from RAO~\SI{90}{\degree} to LAO~\SI{90}{\degree}, simulated fluoroscopic images showed visually imperceptible differences from the real ones, achieving a mean target projection distance error of~\SI{1.55}{\milli\metre}. An \enquote{endoleak} phantom insertion experiment highlighted the effectiveness of simulating multiplanar views with real-time instrument overlays, achieving a mean needle tip error of~\SI{3.42}{mm}.
\textbf{Conclusions:}
\textcolor{black}{Results demonstrated the efficacy of virtual fluoroscopy integrated with magnetic tracking, improving depth perception during navigation. The broad capture range of virtual fluoroscopy showed promise in improving the users’ understanding of X-ray imaging principles, facilitating more efficient image acquisition.}}


%%\pacs[JEL Classification]{D8, H51}

%%\pacs[MSC Classification]{35A01, 65L10, 65L12, 65L20, 65L70}

\maketitle
%\vspace{-0.5cm}
\section{Introduction}\label{intro}
X-ray fluoroscopy is routinely used for intraoperative guidance in cardiovascular, endovascular, orthopedic and neuro-interventions~\cite{merloz2007fluoroscopy, cazzato2020spinal, nijkamp2019prospective}. It provides high-contrast visualization of bones, contrast-enhanced areas, and metallic objects, enabling real-time monitoring of surgical instruments. To accurately assess instrument depth, the C-arm must be oriented perpendicular to the instrument’s trajectory and, after evaluating the depth, reverted to its original pose (\eg, bird's eye view) to monitor the orientation of the instrument. This process often needs to be repeated multiple times to safeguard the precise instrument placement. Additionally, radiologists often require specific fluoroscopic views, such as the pedicle, fracture fragment, or joint space, which are only visible when captured from a limited C-arm range. Achieving the desired radiographic view in practice is a trial-and-error process, known as “fluoro-hunting”, involving frequent C-arm repositioning and repeated image acquisition. These limitations stem from the 2D projective nature of X-ray imaging, resulting in two major drawbacks of fluoroscopy-guided procedures: limited depth perception and prolonged radiation exposure to both patients and medical personnel.

Virtual fluoroscopy~\cite{foley2001virtual} is a promising strategy to mitigate the limitations of conventional fluoroscopy guidance. This technique generates simulated fluoroscopic images at desired C-arm poses using \textcolor{black}{preoperative} CT images, while \textcolor{black}{optionally} overlaying tracked surgical instruments in real time. \textcolor{black}{Specifically, virtual fluoroscopy can 1) improve depth perception. Foley \etal~\cite{foley2001virtual} developed a single C-arm virtual fluoroscopy system to enable real-time, multiplanar procedural guidance. The surgical instrument's movement was updated across multiple fluoroscopic views in real time, without needing C-arm repositioning to reacquire these views. Since there are no pre-acquired 3D images involved in this approach, it cannot provide non-projective views of the anatomy, such as axial, sagittal or coronal. Similarly, Suhm \etal~\cite{suhm2001intraoperative} proposed a virtual fluoroscopy-based technique for distal locking of intramedullary implants, evaluating the intraoperative accuracy using an artificial landmark. 2) reduce radiation exposure.} 
Most current solutions aim to improve users’ understanding of image acquisition principles, thereby facilitating efficient C-arm repositioning during procedures. For example, Navab \etal~\cite{navab2009camera}'s Camera Augmented Mobile C-arm system led to the development of a workflow by Dressel \etal~\cite{dressel2010intraoperative} that used camera-trackable patterns to facilitate registering the X-ray imaging system with the patient, thereby generating virtual fluoroscopic images to assist in C-arm positioning. However, this system used a non-standard C-arm setup with the X-ray source positioned over the patient, potentially increasing radiation exposure. In addition, De Silva \etal~\cite{de2018virtual} introduced \textit{FluoroSim}, a virtual fluoroscopy workflow achieved via C-arm mechanical encoding, geometric calibration, and patient fluoro-CT registration. Their results demonstrated that \textit{FluoroSim} could reduce radiation exposure during C-arm repositioning without compromising positioning time and accuracy.

\textcolor{black}{In} virtual fluoroscopy, integrating spatial trackers into the fluoroscopy-guided workflow is a commonly used strategy. Specifically, tracking sensors are employed to track the surgical instrument, and optionally the C-arm and patient’s anatomy. Magnetic tracking (MT) has gained considerable attention due to its capability to track flexible instruments, such as catheters, guidewires and flexible endoscopes seated subcutaneously~\cite{ramadani2022survey} without the line-of-sight limitation compared to optical trackers. \textcolor{black}{While this capability positions MT as a critical enabler for advanced interventional procedures, its integration is challenging due to the interference from ferromagnetic materials in the C-arm room.} Recent advancements include radiolucent field generators (FGs) that employ error compensation techniques to mitigate the adverse effects of ferromagnetic interference. To the best of our knowledge, our previous work~\cite{xing2024towards} was the first study to integrate a radiolucent tabletop FG prototype into fluoroscopy-guided workflows and demonstrated the efficacy and clinical applicability of our MT-assisted solution in facilitating 3D interventional guidance.

\begin{figure}[h]
\centerline{\includegraphics[width=.9\textwidth]{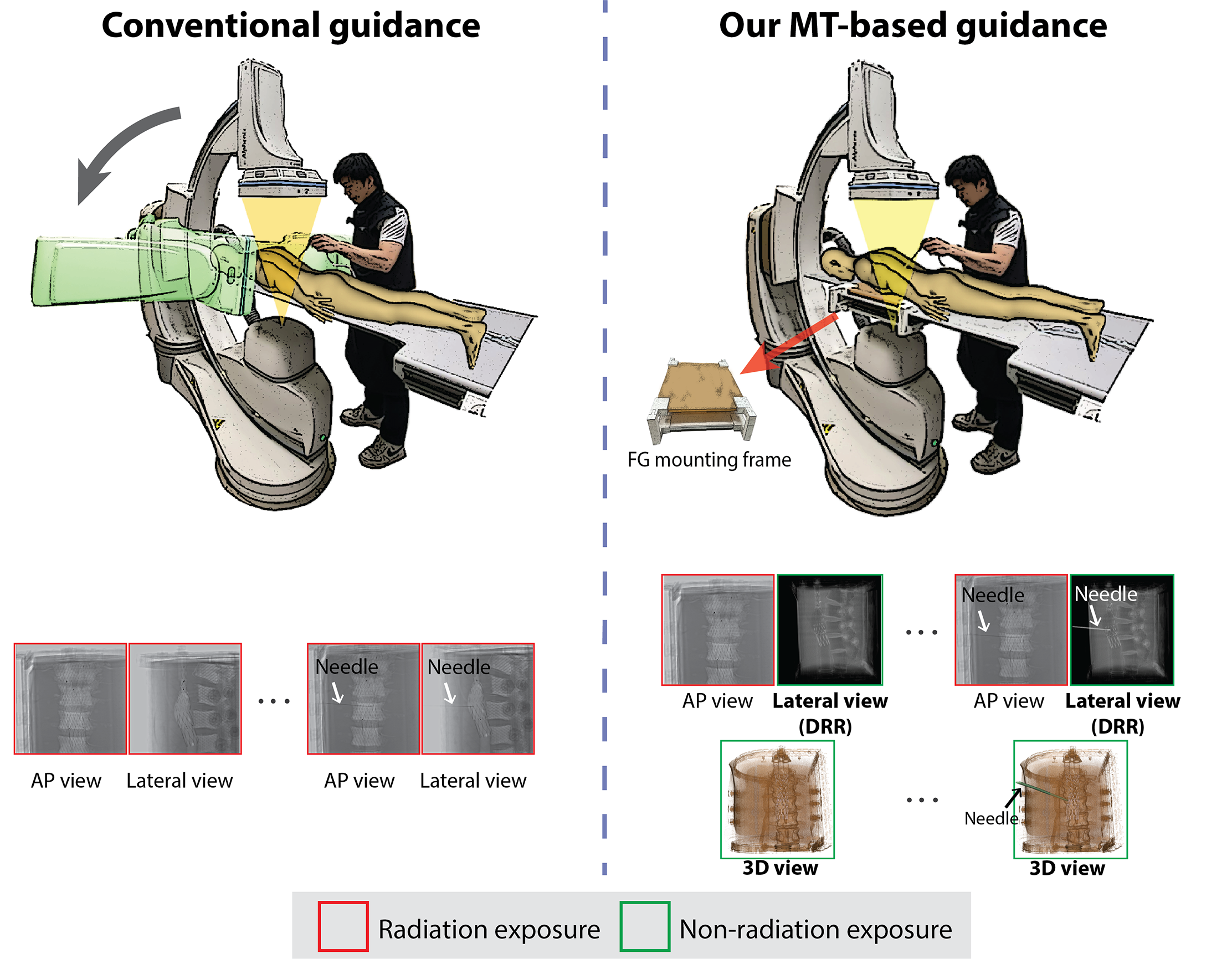}}
\caption{Comparison of our proposed MT-assisted guidance with the conventional one.}
\label{fig_system_overview_comparison}
\end{figure}

%\vspace{-0.5 cm}
%\begin{inparaenum}[1)]
Building upon our previous work, this paper aims to demonstrate the efficacy of incorporating virtual fluoroscopy to enhance depth perception and reduce the number of fluoroscopy acquisitions during procedures. Our contributions are summarized in Fig.~\ref{fig_system_overview_comparison}, including: 

\begin{enumerate}
\renewcommand{\labelenumi}{\theenumi)}

\item We propose an automatic virtual fluoroscopy workflow that leverages our designed two-layer FG mounting frame, integrated with strategically positioned aluminum fiducials. 

\item We present a generalized C-arm modelling approach capable of accurately generating the desired C-arm pose based on either the C-arm’s mechanical encoder data or user-specified viewing parameters. 

\item We develop a 2D-3D landmark correspondence approach to automate the co-registration process between patient fluoroscopic and CT images.
\end{enumerate}
%\end{inparaenum}

\section{Methods}\label{methods}
\subsection{Hardware Integration}\label{hardware_integration}
\begin{figure}[h]
\centering
\begin{tabular}{@{} c @{} c @{} c @{}}
\includegraphics[height=.22\textwidth]{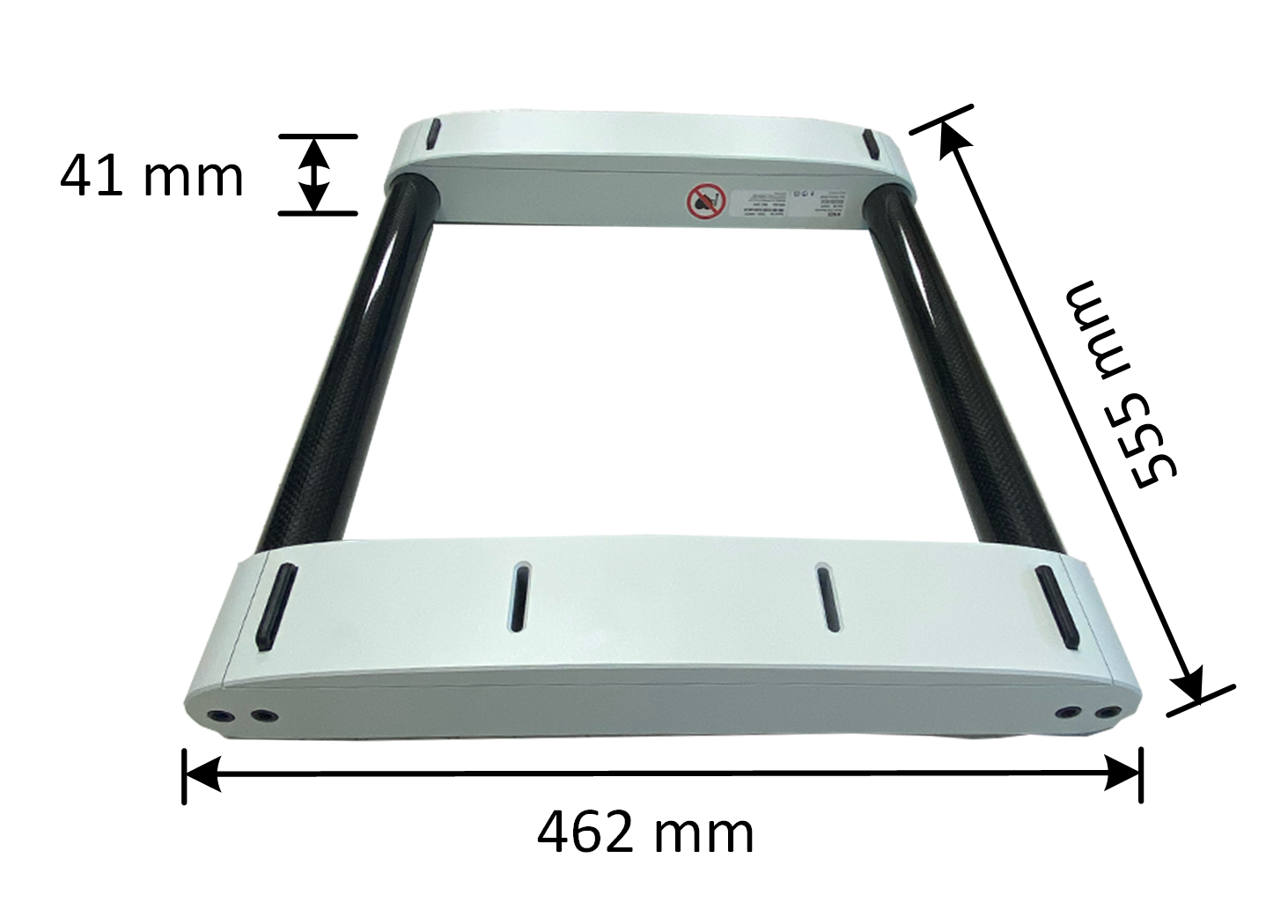}&
\includegraphics[height=.22\textwidth]{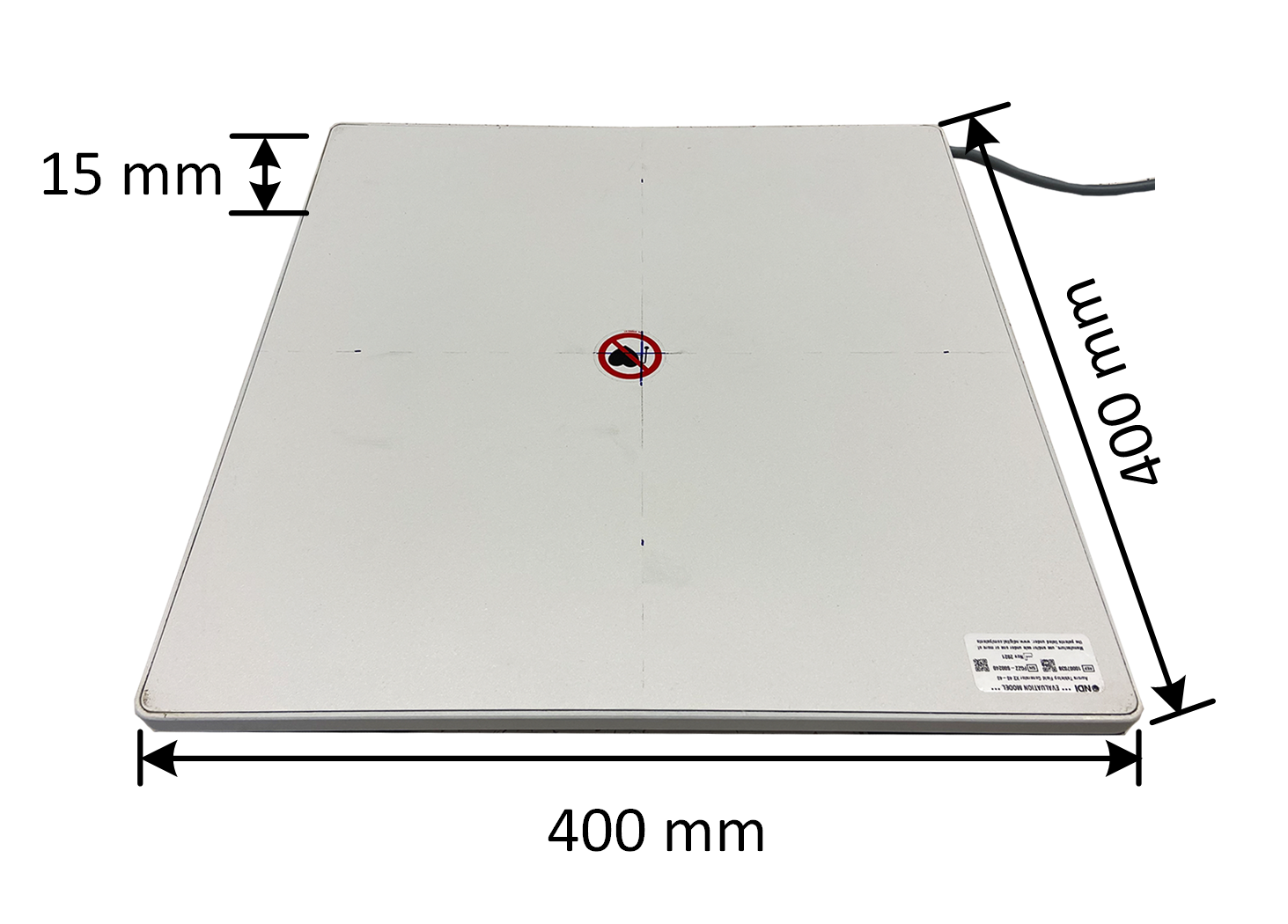}&
\includegraphics[height=.22\textwidth]{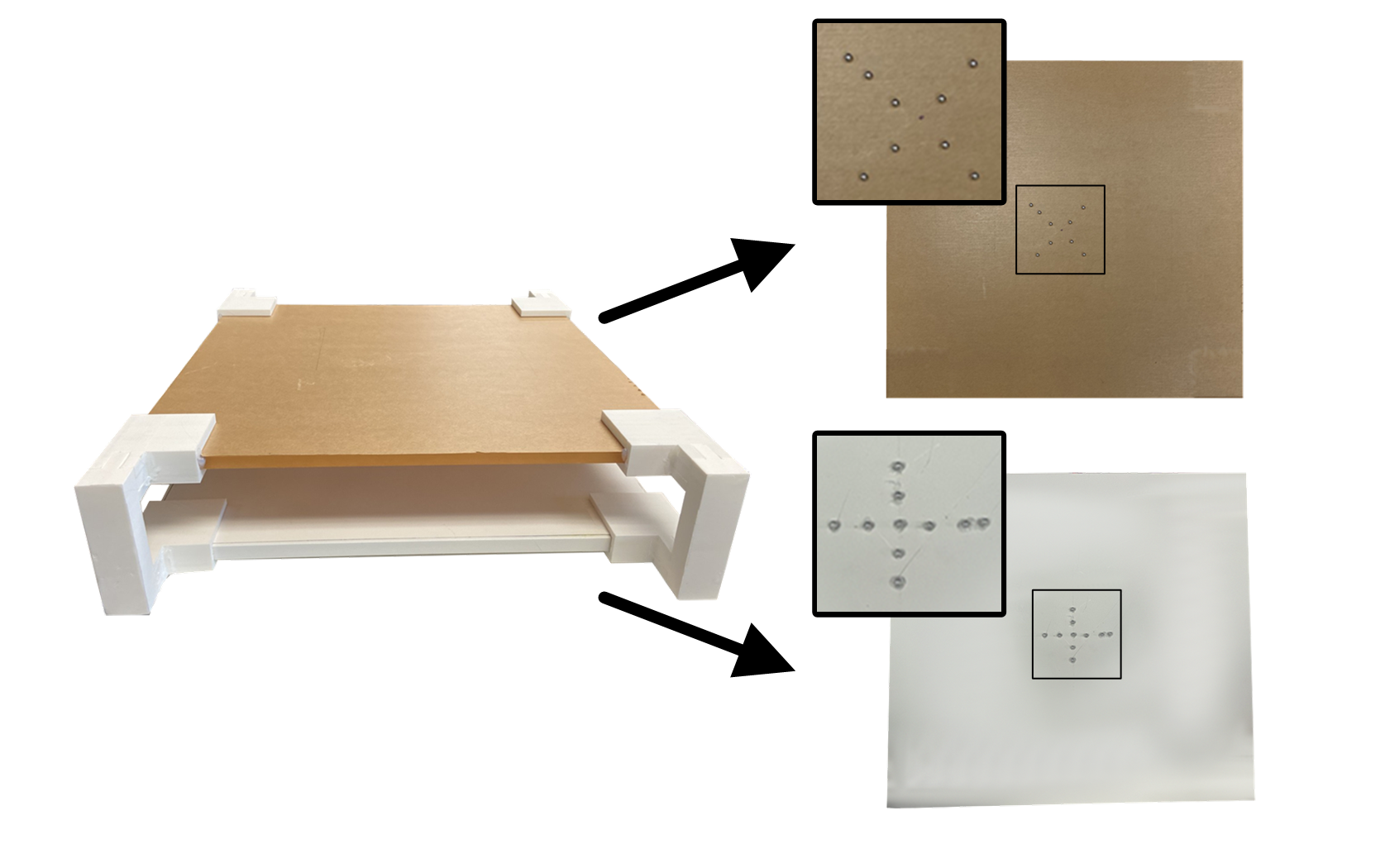}
\\
{(A)} & {(B)} &{(C)}\\
\end{tabular}
\caption{Comparison of the window FG with the radiolucent FG prototype, and the FG mounting frame. The top layer of the FG mounting frame comprises an acrylic plate, and the bottom layer houses the radiolucent FG prototype. Fiducials are attached to the downside of each layer.}
\label{fig_NDI_FG_mounting_frame}
\end{figure}

We employed a magnetic tracking system, featuring a prototype radiolucent FG (TTFG45-55T) (Northern Digital Inc., Ontario, Canada). The unique construction with radiolucent materials results in a reduction of metal-induced imaging artifacts in X-ray radiographs when placed under the path of X-ray beam compared to other conventional field generators. The radiolucent FG prototype has also a thinner profile than the window FG and the form factor of a tabletop design enables easy integration with the surgical bed, as shown in Fig.~\ref{fig_NDI_FG_mounting_frame}. A two-layer FG mounting frame was designed to allow the placement of the radiolucent FG prototype in proximity to the surgical site while maintaining optimal tracking accuracy. Fig.~\ref{fig_system_overview_comparison} depicts the setup of integrating the FG mounting frame into the C-arm suite. Within the mounting frame, nineteen \SI{4}{\milli\metre}-diameter aluminum spherical fiducials were attached to the downside of each layer (Fig.~\ref{fig_NDI_FG_mounting_frame}C) for C-arm pose estimation. \textcolor{black}{The compact FG mounting frame design supports both instrument tracking and fluoro-CT registration (see section~\ref{virtual_fluoroscopy}), offering a space-efficient and functional design well-suited for crowded operations rooms.}

%\vspace{-0.8 cm}
\subsection{Virtual Fluoroscopy}\label{virtual_fluoroscopy}
\begin{figure}[h]
\centering
%\begin{tabular}{@{} c @{}}
\includegraphics[width=.9\textwidth]{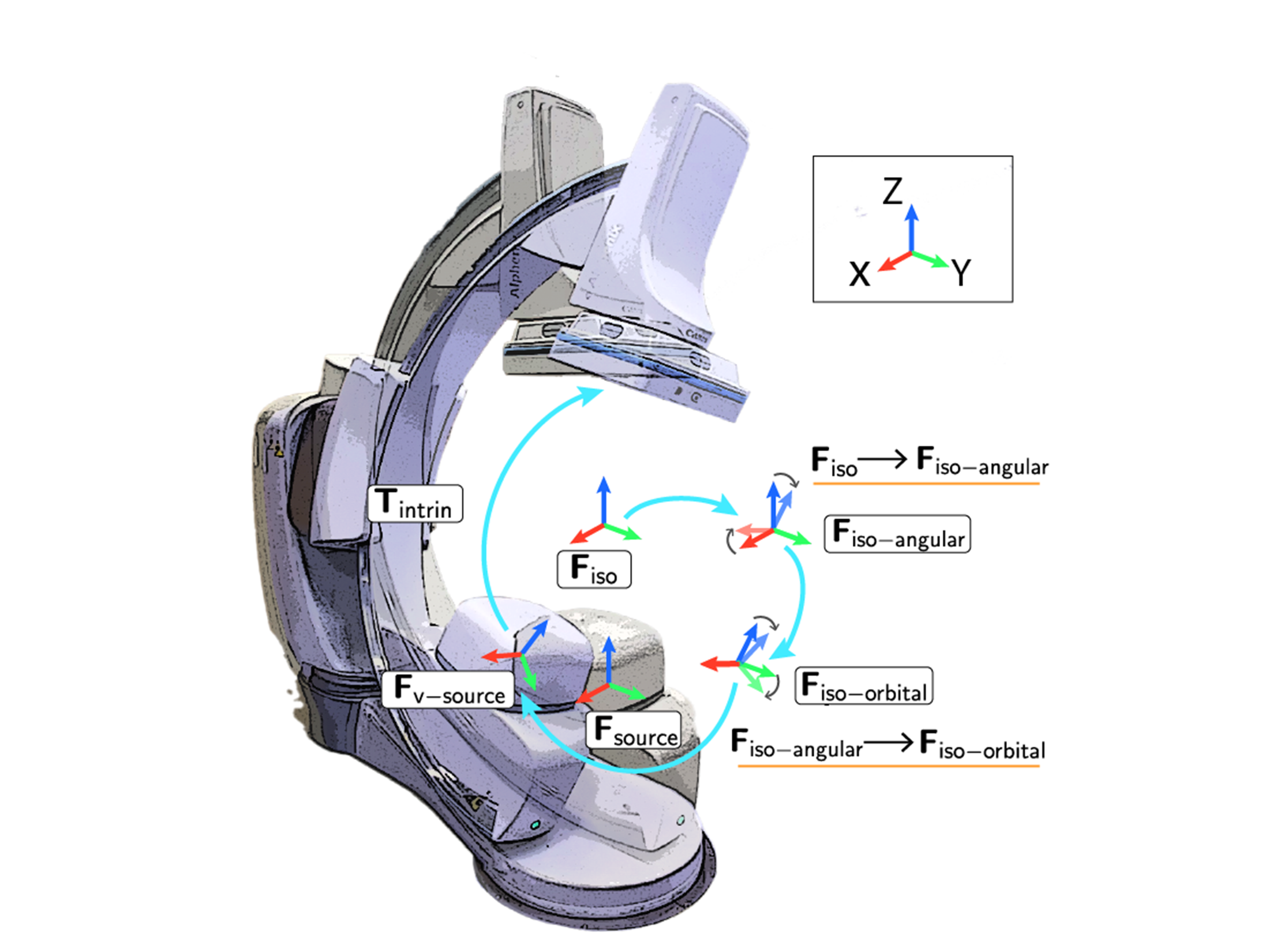}
%\\
%{(A)}\\
%\end{tabular}
\caption{System coordinate frames for virtual fluoroscopy.}
\label{fig_VF_model}
\end{figure}

As depicted in Fig.~\ref{fig_VF_model}, several coordinate frames were defined to elucidate the virtual fluoroscopy workflow, including the physical X-ray source frame ($F_{source}$), the virtual X-ray source frame ($F_{v-source}$), \textcolor{black}{the coordinate frame of the patient CT image ($F_{patient-CT}$),} the C-arm rotation center frame ($F_{iso}$), the coordinate frame ($F_{iso-angular}$) that is defined after angular rotation along the left-anterior-oblique (LAO) to right-anterior-oblique (RAO) direction relative to $F_{iso}$, and the coordinate frame ($F_{iso-orbital}$) that is defined after orbital rotation along the cranial (CRA) to caudal (CAU) direction relative to $F_{iso-angular}$. The X, Y and Z axes of $F_{iso}$ (Fig.~\ref{fig_VF_model}) align with the C-arm's left-right, cranial-caudal, and up-down directions, respectively. In $F_{source}$, the Z axis points from the X-ray source to the image detector, the Y axis follows the cranial-caudal direction, and the X axis is determined by the cross product of the Y and Z axes. The mathematical model of C-arm virtual fluoroscopy is described by Eqn.~\ref{eqn1}:
\begin{equation}\label{eqn1}
    I_{v-fluoro} = \mathcal{DRR}(I_{\textcolor{black}{patient-CT}}, T_{intrin}, T_{\textcolor{black}{patient-CT}}^{v-source})
\end{equation} 

\noindent where the virtual fluoroscopic image ($I_{v-fluoro}$) is generated by ray-casting the patient CT image ($I_{\textcolor{black}{patient-CT}}$) using the C-arm intrinsic matrix ($T_{intrin}$) and the desired C-arm pose ($T_{\textcolor{black}{patient-CT}}^{v-source}$). \textcolor{black}{$I_{patient-CT}$ accurately represents the real patient's anatomical structures assuming that only rigid motion within the region of interest.} Note that in this context, $T_A^B$ denotes a 4$\times$4 transformation matrix from frame $F_A$ to frame $F_B$.
\begin{equation}\label{eqn2}
    T_{\textcolor{black}{patient-CT}}^{v-source} = T_{iso-orbital}^{v-source} * T_{iso-angular}^{iso-orbital}(\beta) * T_{iso}^{iso-angular}(\alpha) * T_{\textcolor{black}{patient-CT}}^{iso} 
\end{equation}
The C-arm intrinsic parameters, including pixel spacing, image size and focal length, can be derived from the system settings. Eqn.~\ref{eqn2} defines the transformation chain used to calculate the only unknown $T_{\textcolor{black}{patient-CT}}^{v-source}$. The angular rotation ($\alpha$), orbital rotation ($\beta$), and ISO distance ($T_{iso-orbital}^{v-source}$) are obtained either from the C-arm's mechanical encoding or the user-defined view (\eg, lateral view). Thus, $T_{\textcolor{black}{patient-CT}}^{iso}$, representing the transformation from the patient \textcolor{black}{CT} to the C-arm rotation center ($F_{iso}$), remains the only unknown.
\begin{equation}\label{eqn3}
    T_{\textcolor{black}{patient-CT}}^{iso} = T_{\textcolor{black}{patient-CT}}^{source} * (T_{iso-orbital}^{source} * T_{iso-angular}^{iso-orbital}(\beta_0) * T_{iso}^{iso-angular}(\alpha_0))^{-1} 
\end{equation}

To compute $T_{\textcolor{black}{patient-CT}}^{iso}$, an initial fluoroscopic image is acquired along with the corresponding angular ($\alpha_0$) and orbital ($\beta_0$) rotations. A fluoro-CT registration approach is then applied to calculate $T_{\textcolor{black}{patient-CT}}^{source}$ (see below), from which $T_{\textcolor{black}{patient-CT}}^{iso}$ can be derived as outlined in Eqn.~\ref{eqn3}, assuming patient remains stationary during the image acquisition. The task to calculate $T_{\textcolor{black}{patient-CT}}^{source}$ essentially is the classic fluoro-CT registration problem combined with the known C-arm intrinsic parameters. 
\subsection{Fluoro-CT Registration}

\begin{figure}[h]
\centering
%\begin{tabular}{c}
\includegraphics[width=.98\textwidth]{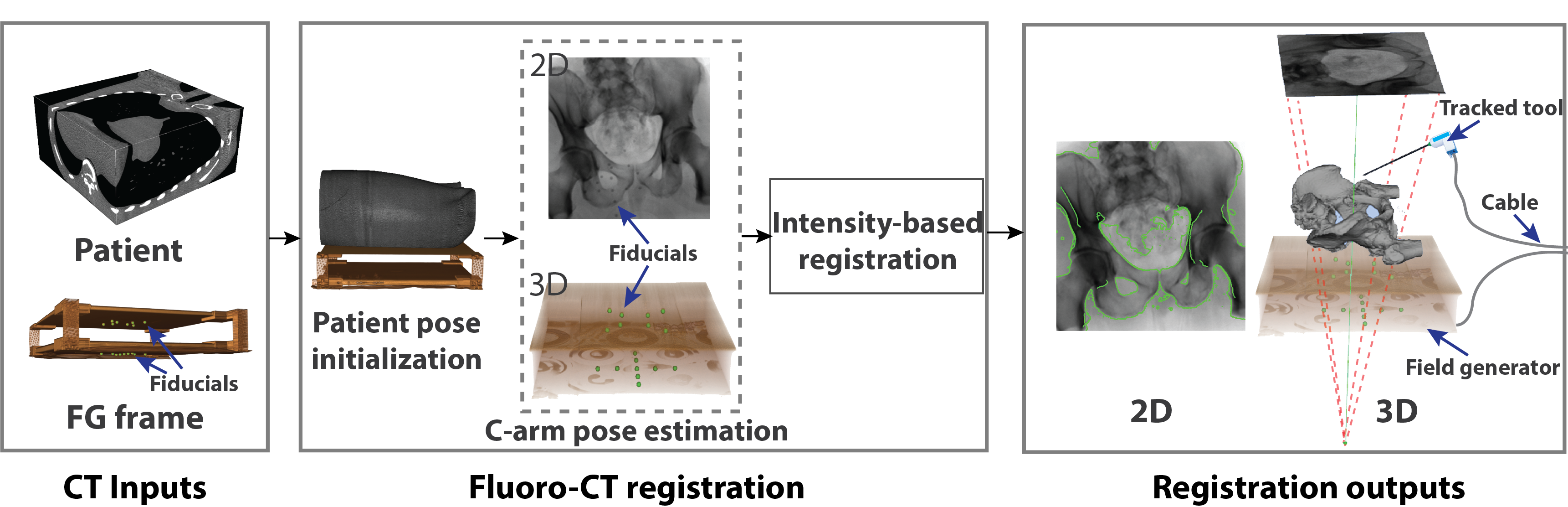}
%{(A)}\\
%\end{tabular}
\caption{\textcolor{black}{Workflow of fluoro-CT registration.}}
\label{fig_fluoro_CT_registration}
\end{figure}

As detailed in \textcolor{black}{\cite{xing2024towards}}, \textcolor{black}{the} registration approach \textcolor{black}{includes} three steps: patient pose initialization, C-arm pose estimation, and intensity-based registration\textcolor{black}{, as shown in Fig.~\ref{fig_fluoro_CT_registration}}. \textcolor{black}{The required inputs include a patient CT image and a CT image of the FG mounting frame, which contains the attached 3D fiducials, as shown in the CT inputs in Fig.~\ref{fig_fluoro_CT_registration}. Notably, these two CT images (the patient and the FG mounting frame) are independent and not co-registered as the input stage.}

%\vspace{-0.5 cm}
\subsubsection{Patient Pose Initialization}
%\vspace{-0.3 cm}
This step entails positioning the patient’s region of interest on the FG mounting frame according to clinical requirements. Specifically, after patient pose initialization, the relative relationship between $I_{\textcolor{black}{patient-CT}}$ and \textcolor{black}{the CT image of} the FG mounting frame should approximate the actual clinical setup. It is worth noting that this step has a high tolerance of misalignment. As long as the patient CT is approximately aligned, accurate fluoro-CT registration can be achieved using the following steps.

\subsubsection{C-arm Pose Estimation}
\begin{figure}[h]
\centering
\begin{tabular}{c}
\includegraphics[width=.90\textwidth]{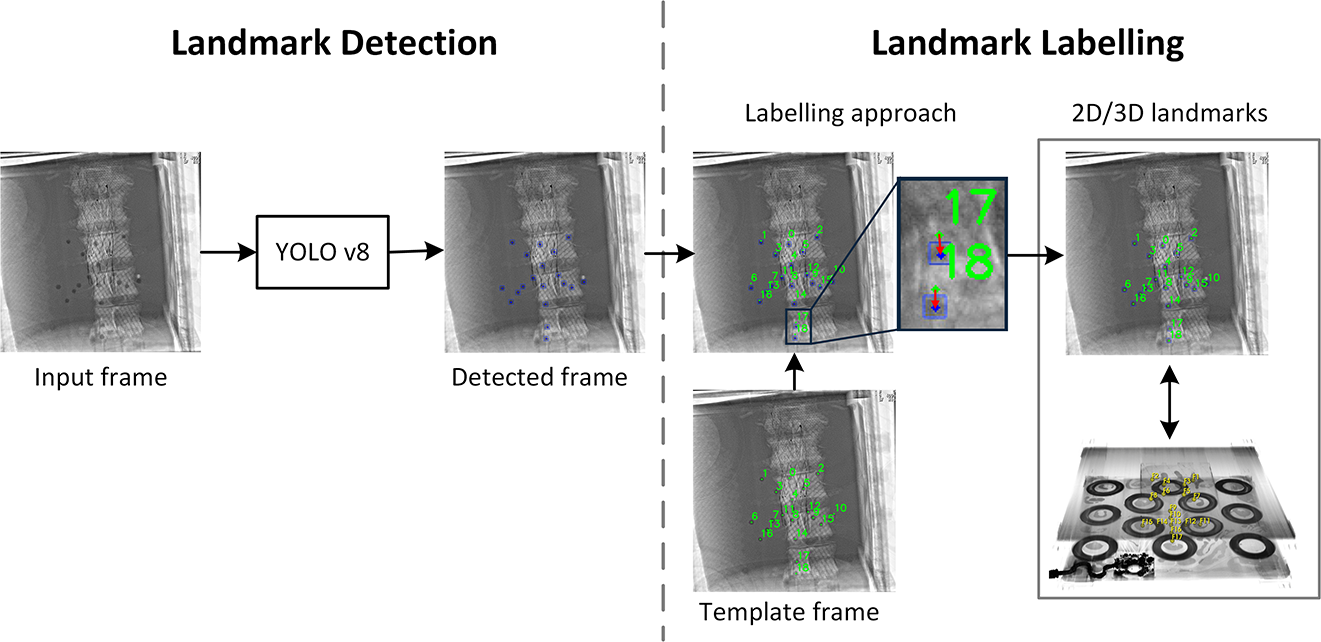}
\\
%{(A)}\\
\end{tabular}
\caption{Workflow of our 2D-3D landmark correspondence approach.}
\label{fig_landmark_correspondence}
\end{figure}

The C-arm pose, defined as the FG mounting frame's pose relative to $F_{source}$, was calculated using a Perspective-n-Point (PnP) approach~\cite{hartley2003multiple}. Aluminum fiducials attached to the FG mounting frame were used to establish the 2D-3D shared landmark correspondence. Additionally, we proposed an \textcolor{black}{automatic} 2D-3D landmark correspondence approach for automatic matching. As shown in Fig.~\ref{fig_landmark_correspondence}, our approach comprises two steps: landmark detection and labelling.

%\vspace{-0.5 cm}
\textit{Landmark Detection}. The You Only Look Once (YOLOv8)\footnote{https://yolov8.com/} object detection model was used for automatic detection of fiducial landmarks in fluoroscopic images. The model was trained on a dataset of 133 fluoroscopic images of the FG mounting frame and an \enquote{endoleak} phantom placed on top. The \enquote{endoleak} phantom contains spinal elements, an aortic stent, a simulated aneurysm and feeding vessels, embedded in polyvinyl alcohol cryogel (PVA-C) as a tissue mimic. Ground truth was generated by manually annotating the fiducial centroids using 3D Slicer software~\footnote{https://www.slicer.org/}. Thirty validation images were used to train the model over 100 epochs. The dataset incorporated variations in C-arm settings, including angular and orbital rotations, source-to-detector (SID) distances, and phantom movement to simulate patient motion, accommodating typical intraoperative changes in C-arm pose and patient positioning.

\textit{Landmark Labelling}. To establish \textcolor{black}{2D-3D} landmark correspondence, consistent labelling of markers across fluoroscopic frames is required as the C-arm moves. We assume that fluoroscopic image frames are being captured in conjunction with the continuous motion of the C-arm along its rotational axes (LAO-RAO, CRA-CAU) and varying SID changes. Using a \enquote{template} of 19 labelled landmark centroids from an initial C-arm pose, we track each centroid to the next frame by minimizing the Euclidean distance between labelled centroids and the set of centroids as detected by the YOLO model. The initial labelled centroids are replaced by their corresponding closest centroid, and the YOLOv8 detection and distance-minimization process is repeated for the subsequent frame.

\subsubsection{Intensity-based Registration}
Beginning with the initial C-arm pose, a multi-resolution image intensity-based registration strategy~\cite{powell2009bobyqa, van2005standardized}, was employed to optimize the pose parameters (\ie, $T_{\textcolor{black}{patient-CT}}^{source}$). This was achieved by employing similarity metrics that compare simulated fluoroscopic images, derived from digitally reconstructed radiographs (DRRs), with the real fluoroscopic image. Due to variations in peak kilovolt (kVp) settings between $I_{\textcolor{black}{patient-CT}}$ and fluoroscopic images, a patch-based gradient normalized cross correlation~\cite{powell2009bobyqa} metric was employed as the objective function for optimization.

\subsection{Evaluation Metrics}
The accuracy of both the fluoro-CT registration and virtual fluoroscopy was evaluated using a 2D metric, mean projection distance (mPD). The mPD measures the distance between the 2D projections of target points at the registration/simulated position, and the projections of target points at the ground-truth position, which can be defined as: 
\begin{equation}\label{eq5}
mPD = \frac{1}{N} \sum_{i=1}^{N} \| T_{intrin}*\dot{T}_{patient}^{source} * P_{i} - T_{intrin}*T_{patient}^{source} * P_{i} \|
\end{equation} 
The target points in the patient CT space are denoted as $P_{i} (i=1,2,…,n)$. The ground-truth and calculated extrinsics by the fluoro-CT registration or virtual fluoroscopy are represented by $\dot{T}_{patient}^{source}$ and $T_{patient}^{source}$, respectively. As shown in the \SI{0}{\degree} view of Fig.~\ref{fig_virtual_fluoroscopy}, the stent intersection points are used as the target points.

\section{Experiments and Results}\label{experiments_results}
\subsection{\textcolor{black}{Landmark Correspondence and} Fluoro-CT Registration}

As shown in Table~\ref{tab_registration_error}, three fluoroscopic videos, including RAO-LAO (133 frames), CRA-CAU (107 frames) and SID changes (61 frames), were acquired using the \enquote{endoleak} phantom to evaluate the accuracy of the 2D-3D landmark correspondence approach and fluoro-CT registration. For landmark detection, the mean Euclidean distance (mED) between the detected and ground-truth landmarks were $\sim$\SI{0.35}{\milli\metre} across all C-arm settings. The YOLOv8 model yielded success rates (SRs) of 71.11\%, 50\% and 91.67\% for the frame that detected all landmarks, and all frames have at least 80\% of landmarks detected across all settings. In addition, at least 80\% of landmarks were successfully labelled in all frames, enabling robust C-arm pose estimation using a PnP approach. Additionally, 54, 64, and 38 frames from the RAO-LAO, CRA-CAU, and SID settings, respectively, were randomly selected to evaluate fluoro-CT registration, yielding an mPD of less than \SI{0.6}{\milli\metre}.

\begin{table}[htbp]
%\vspace{-1 cm}
\caption{Results of landmark detection, labelling and fluoro-CT registration across various clinical C-arm settings. (SR(100\%): success rate of the frame that all landmarks are detected/labelled, SR(80\%): success rate of the frame that at least 80\% of landmarks are detected/labelled, mED: mean Euclidean distance)}
\centering
%\setlength{\tabcolsep}{5pt}
%\resizebox{0.8\textwidth}{!}{
\begin{tabular}{c |c c c| c c | c}
%\Xhline{2\arrayrulewidth}\
\hline
\multirow{3}{*}{C-arm settings}& \multicolumn{3}{c|}{Landmark detection}&\multicolumn{2}{c|}{Landmark labelling} &\multicolumn{1}{c}{Fluoro-CT registration} \\ 
{}& mED & SR & SR & SR & SR & \multirow{2}{*}{mPD($mm$)}\\
{} & {($mm$)} & {(100\%)} & {(80\%)} & {(100\%)} & {(80\%)} &  \\
\hline

{RAO-LAO} & \multirow{2}{*}{0.35 $\pm$ 0.07} &\multirow{2}{*}{71.11\%} & \multirow{2}{*}{100\%} & \multirow{2}{*}{85.34\%} & \multirow{2}{*}{100\%} & \multirow{2}{*}{0.56 $\pm$ 0.05}\\
{(-30$^{\circ}$ - 30$^{\circ}$)} &&&&&&\\
\hline

{CRA-CAU} & \multirow{2}{*}{0.36 $\pm$ 0.07} & \multirow{2}{*}{50\%} &\multirow{2}{*}{100\%} & \multirow{2}{*}{76.64\%} & \multirow{2}{*}{100\%} & \multirow{2}{*}{0.52 $\pm$ 0.05} \\
{(-30$^{\circ}$ - 30$^{\circ}$)} &&&&&&\\
\hline

{SID} & \multirow{2}{*}{0.36 $\pm$ 0.09} & \multirow{2}{*}{91.67\%} &\multirow{2}{*}{100\%} & \multirow{2}{*}{100\%} & \multirow{2}{*}{100\%} & \multirow{2}{*}{0.59 $\pm$ 0.04}\\
{(90 cm - 125 cm)} &&&&&&\\
%\Xhline{2\arrayrulewidth}
\hline
\end{tabular}%}
\label{tab_registration_error}
\end{table}

\subsection{Virtual Fluoroscopy}

\begin{figure}[htbp]
\centering
\resizebox{0.98\textwidth}{!}{
\begin{tabular}{@{} c @{}  c @{}  c @{}  c @{} c @{}}

%\hline
\includegraphics[width=.19\textwidth]{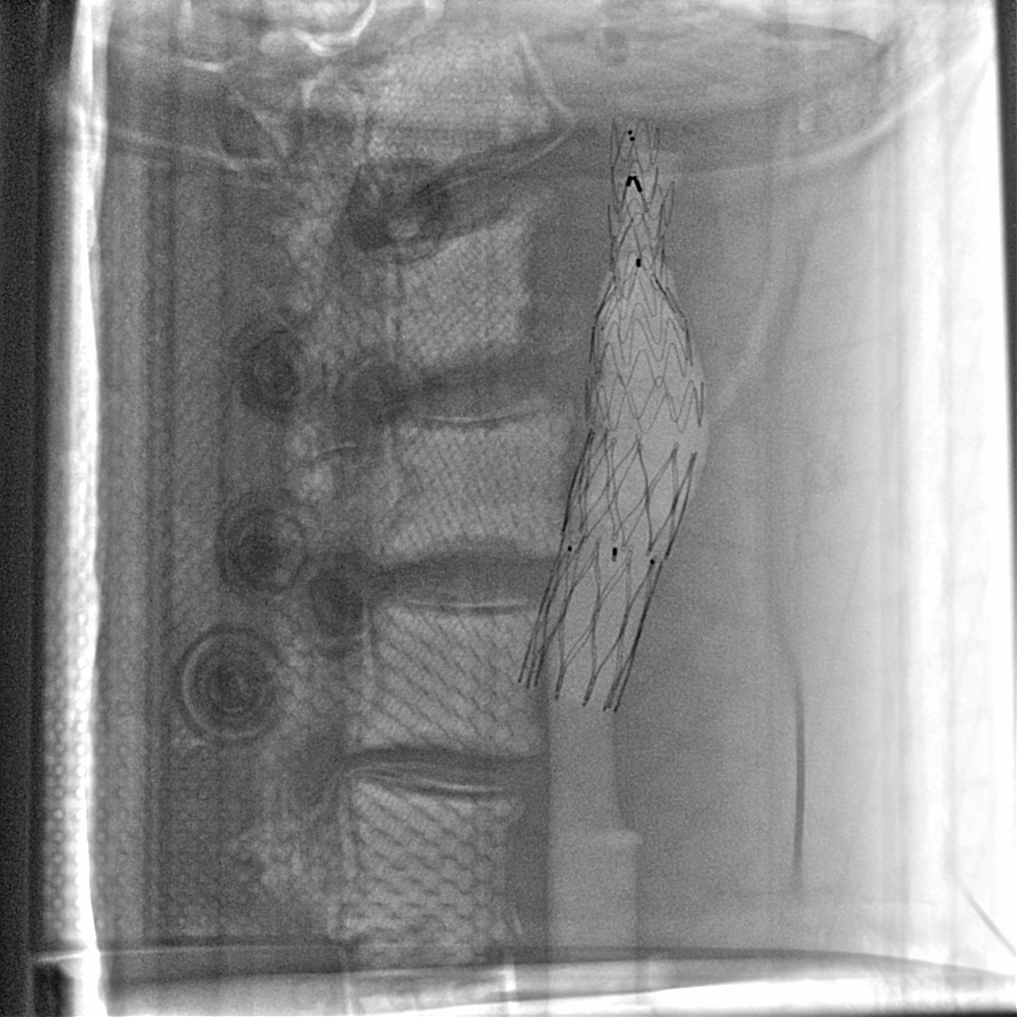}&
\includegraphics[width=.19\textwidth]{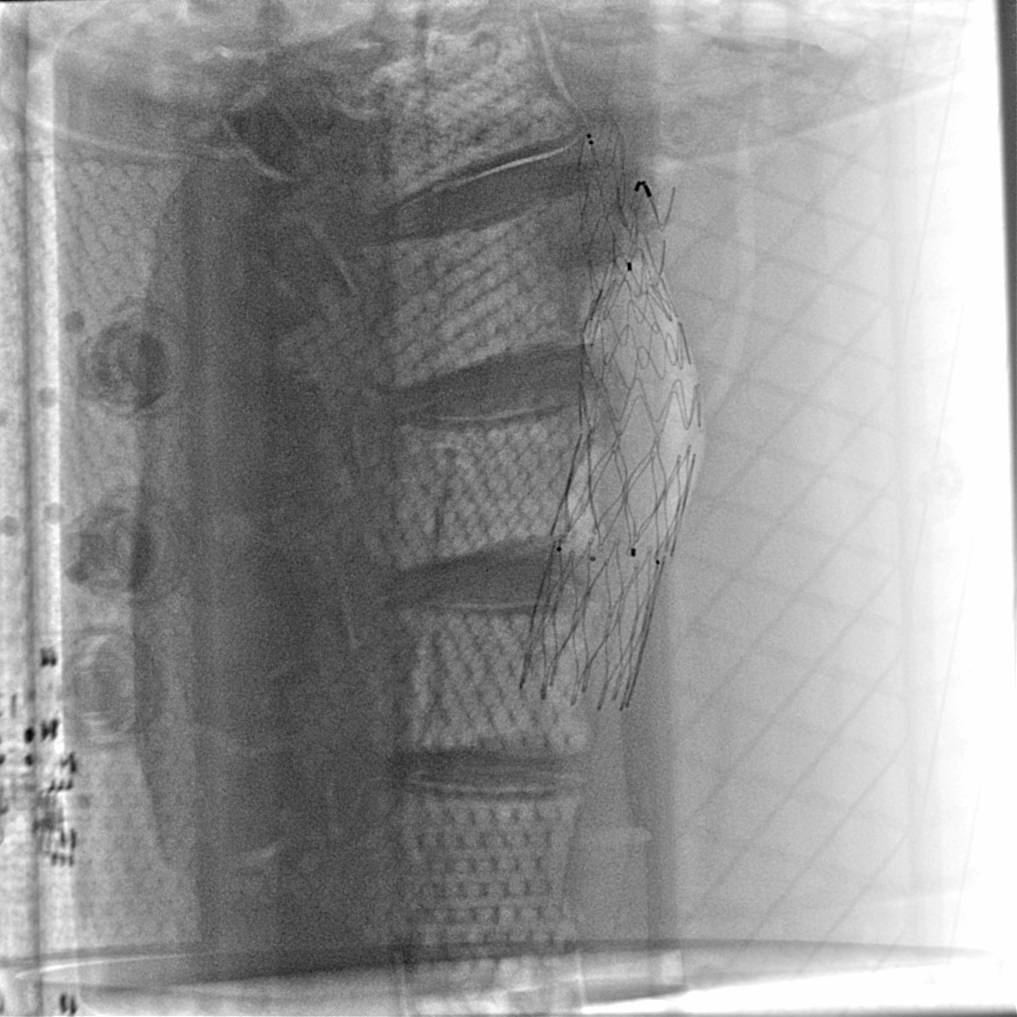}&
\includegraphics[height=.19\textwidth]{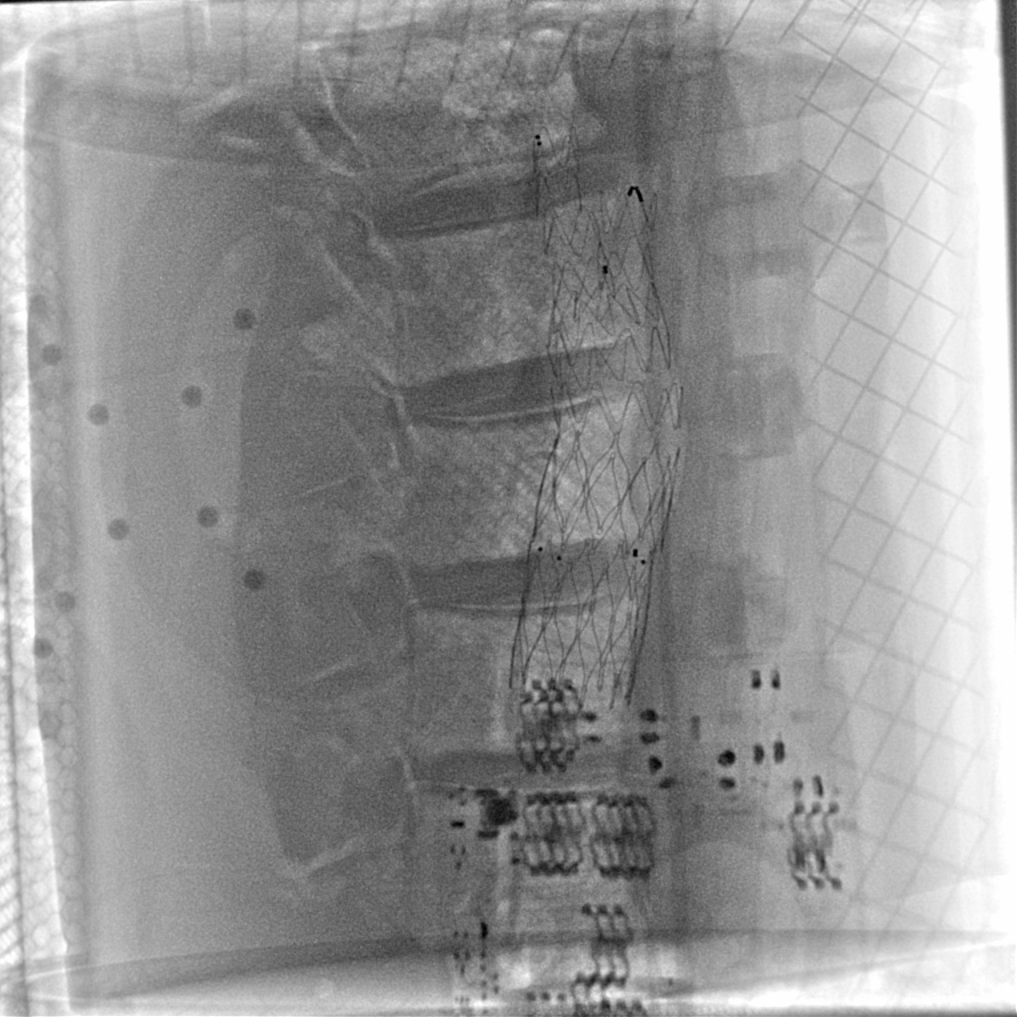}&
\includegraphics[height=.19\textwidth]{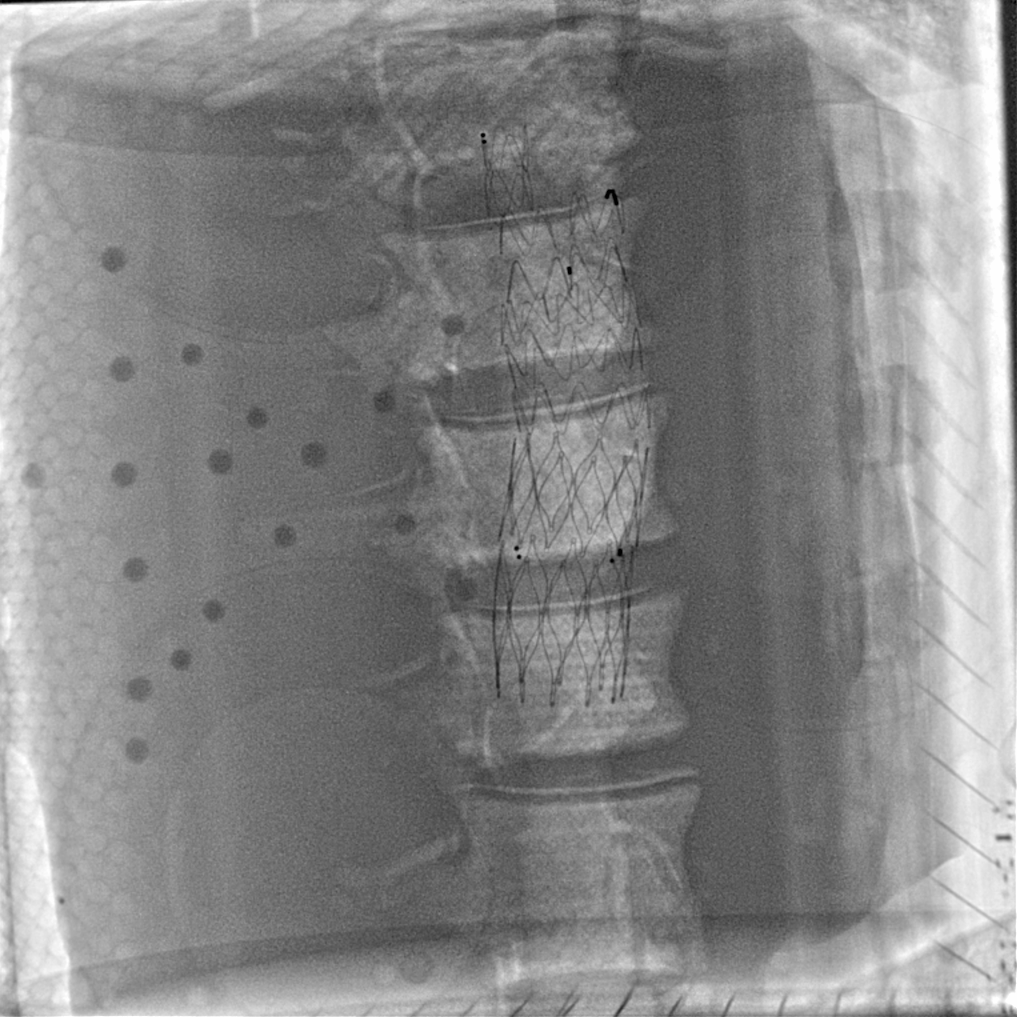}&
\includegraphics[height=.19\textwidth]{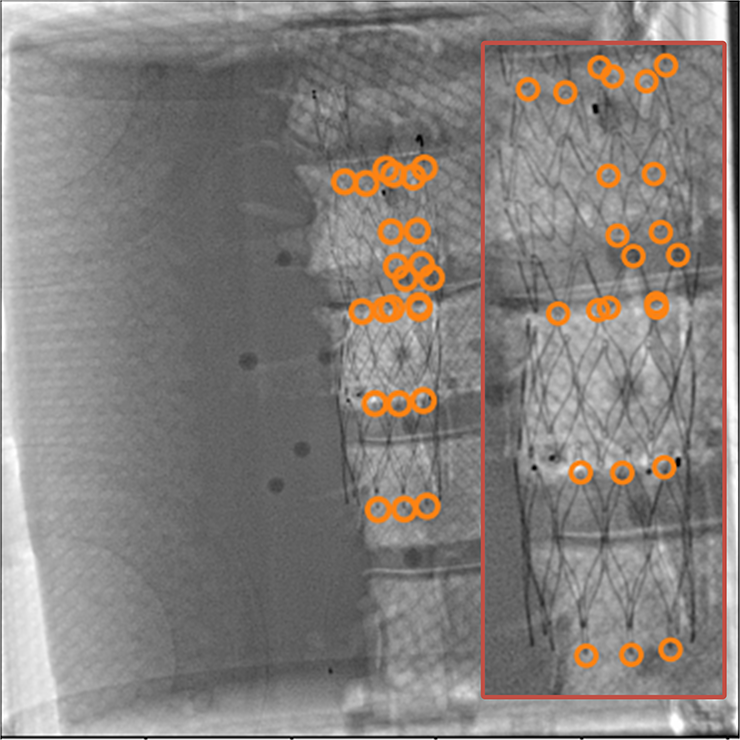}
\\
\includegraphics[width=.19\textwidth]{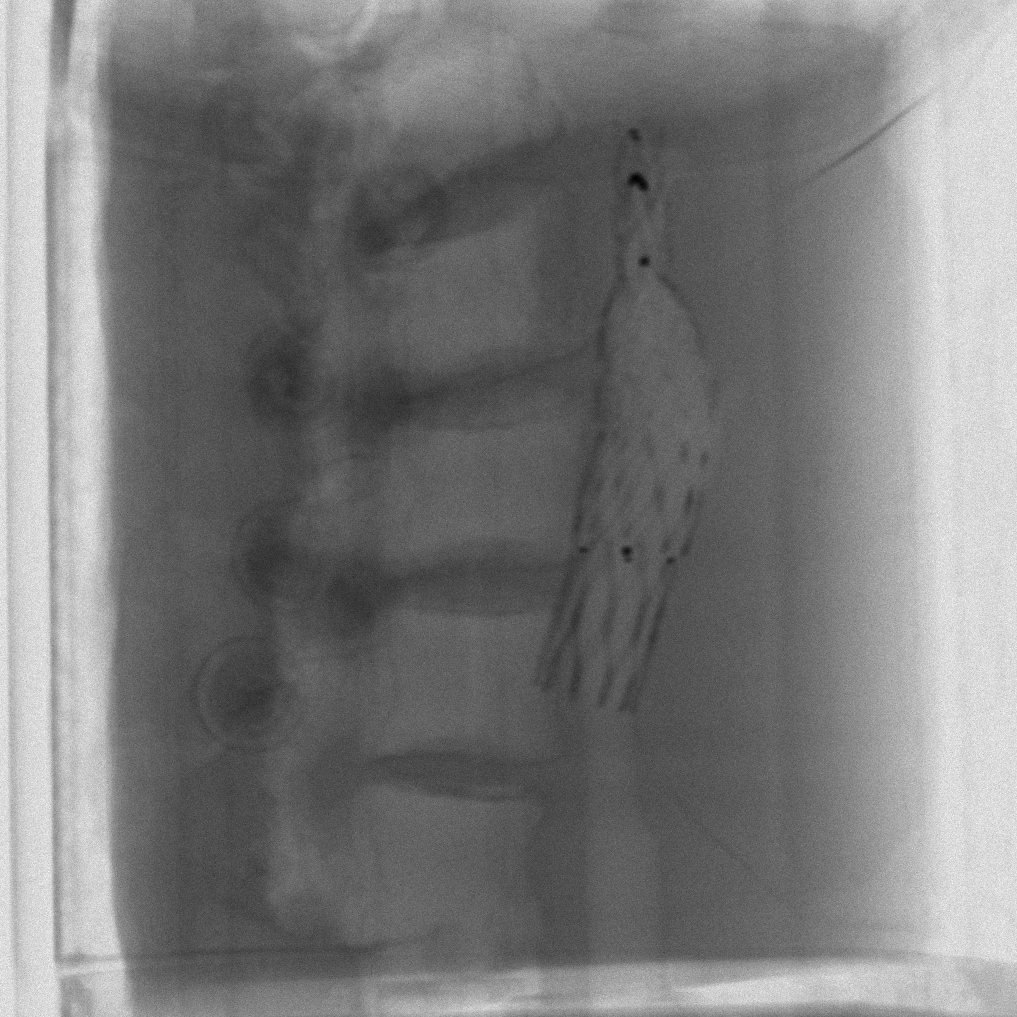}&
\includegraphics[width=.19\textwidth]{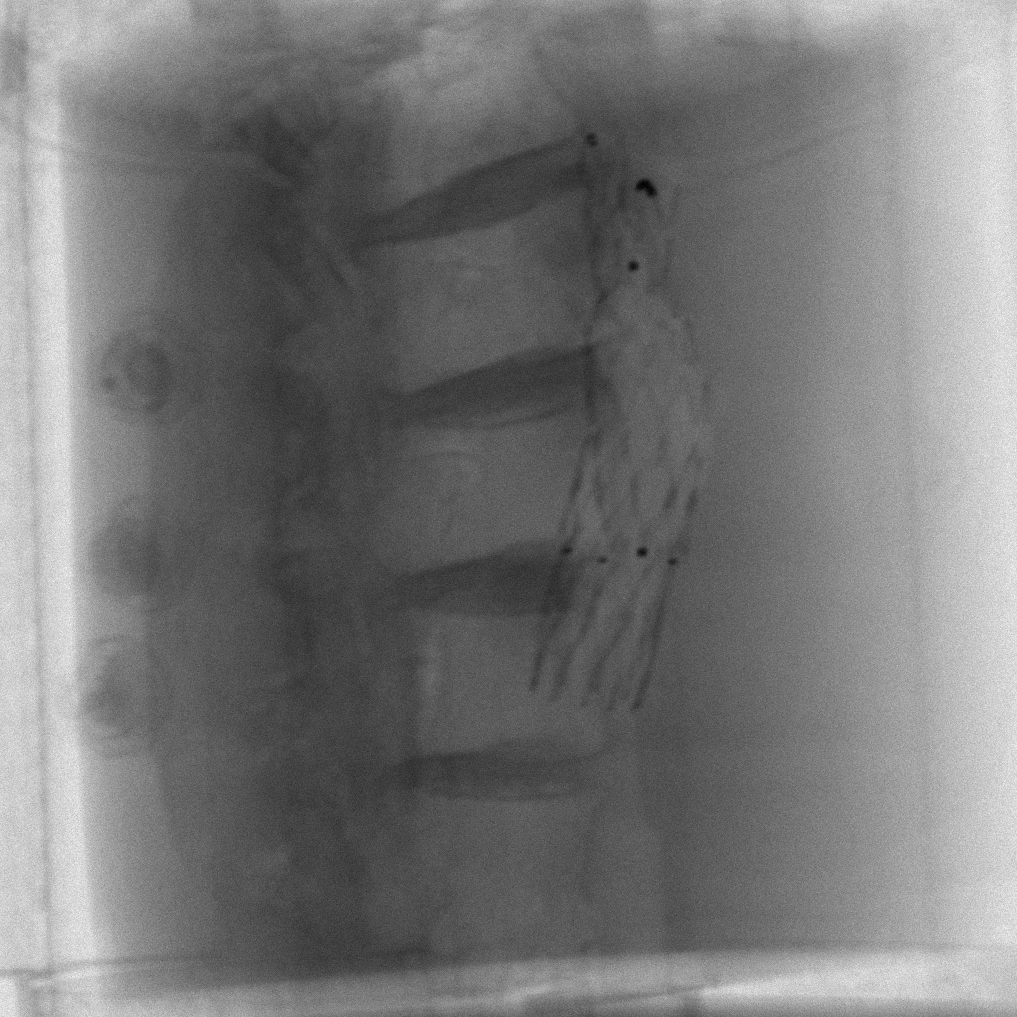}&
\includegraphics[height=.19\textwidth]{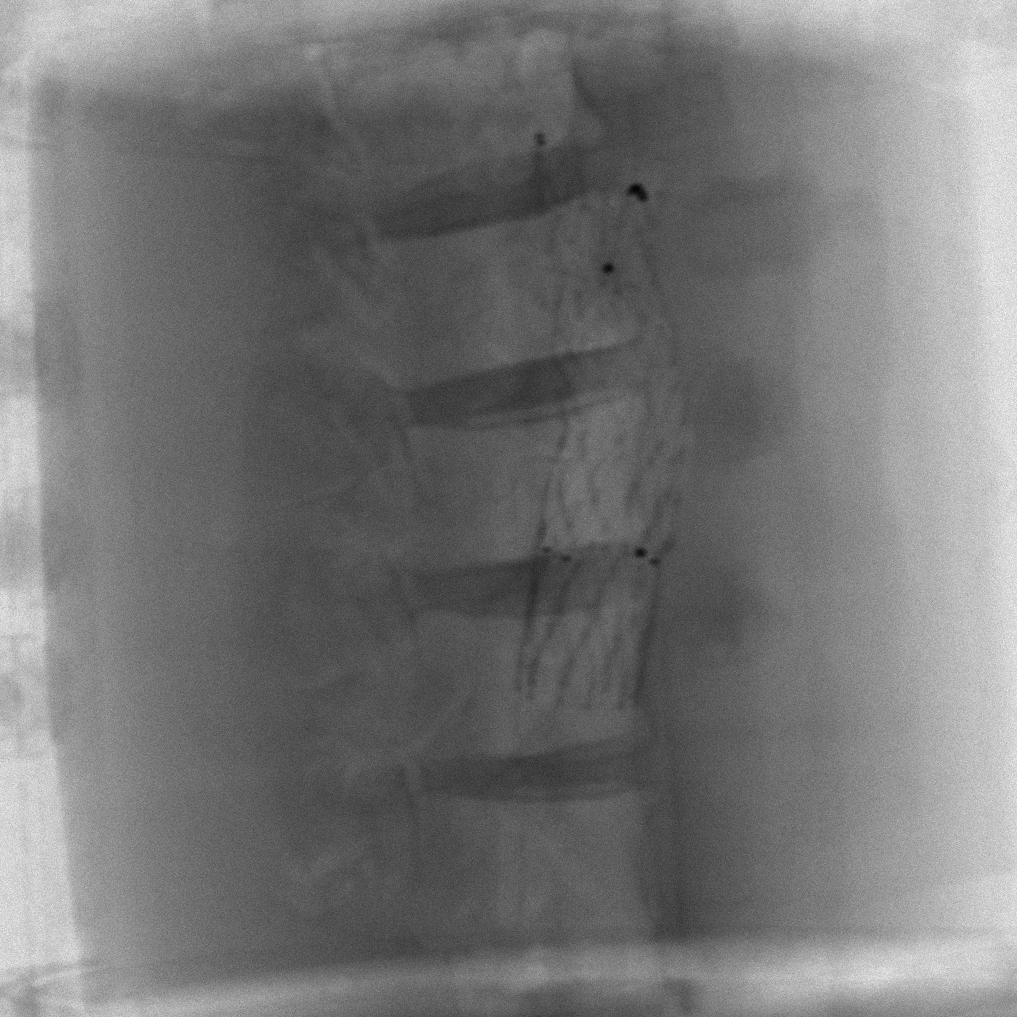}&
\includegraphics[height=.19\textwidth]{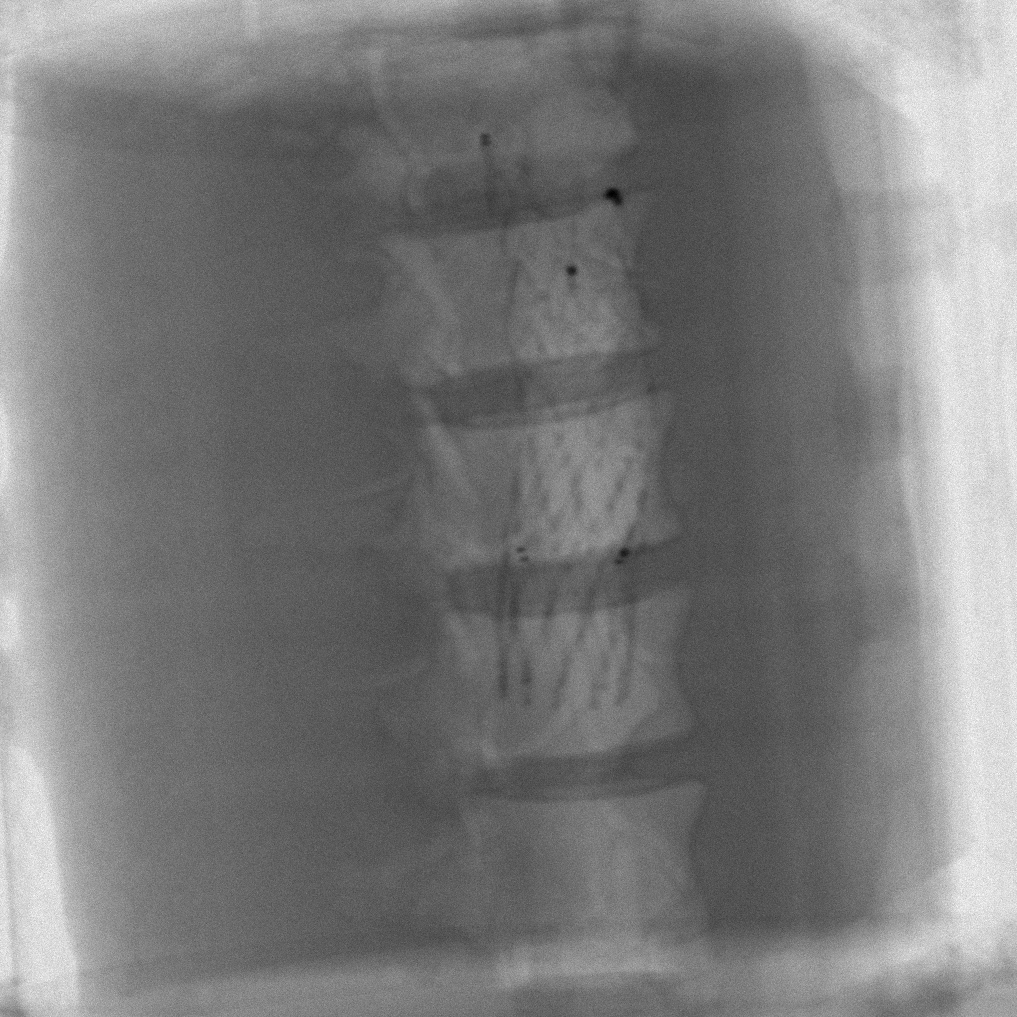}&
\includegraphics[height=.19\textwidth]{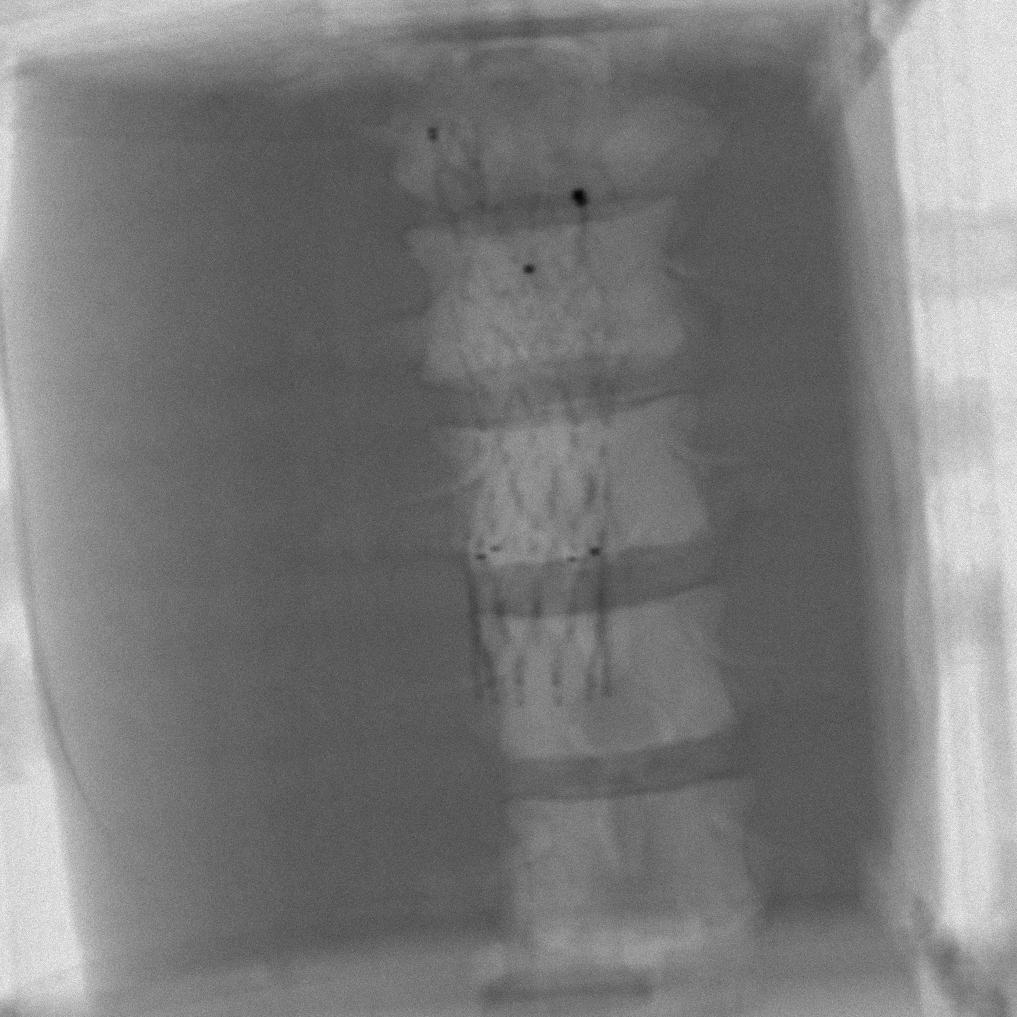}
\\
{RAO~\SI{80}{\degree}} & {RAO~\SI{60}{\degree}} & {RAO~\SI{40}{\degree}} & {RAO~\SI{20}{\degree}} & {\SI{0}{\degree}}
\\
\includegraphics[width=.19\textwidth]{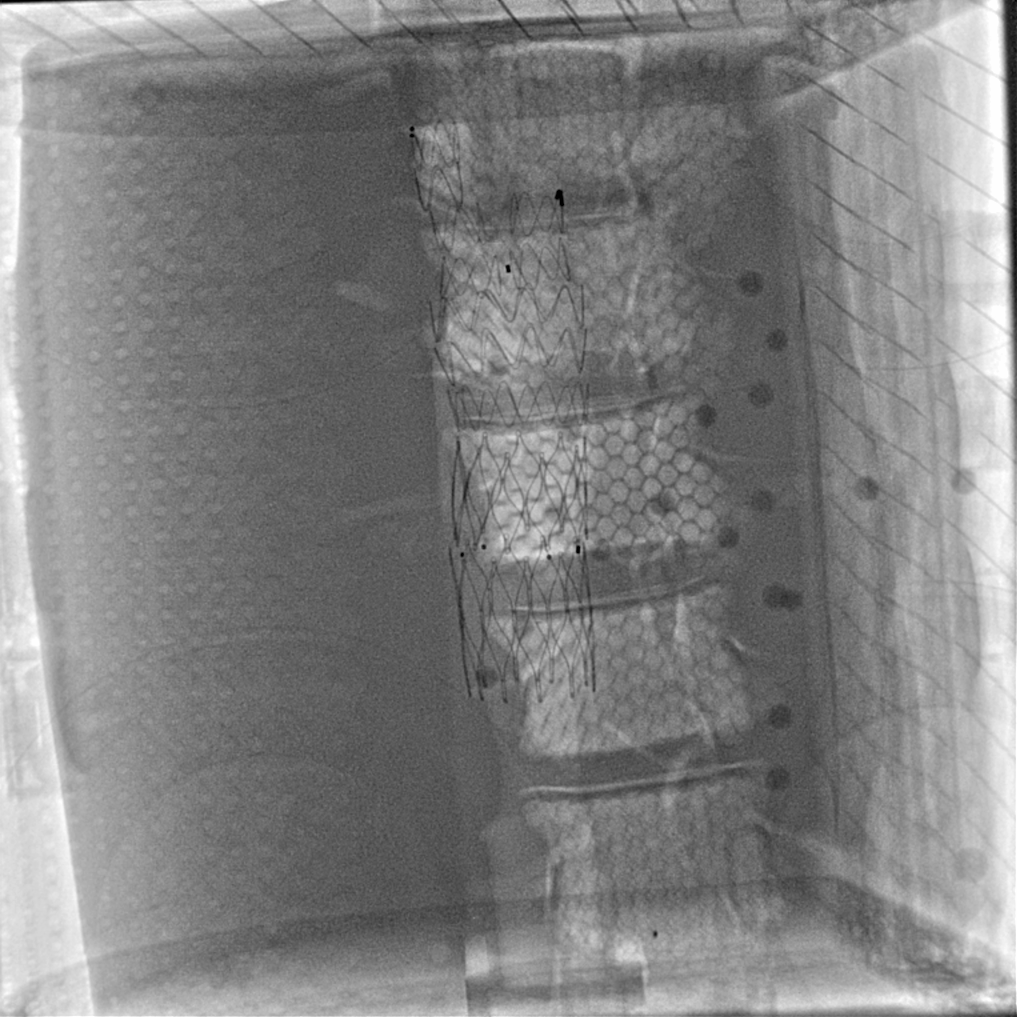}&
\includegraphics[width=.19\textwidth]{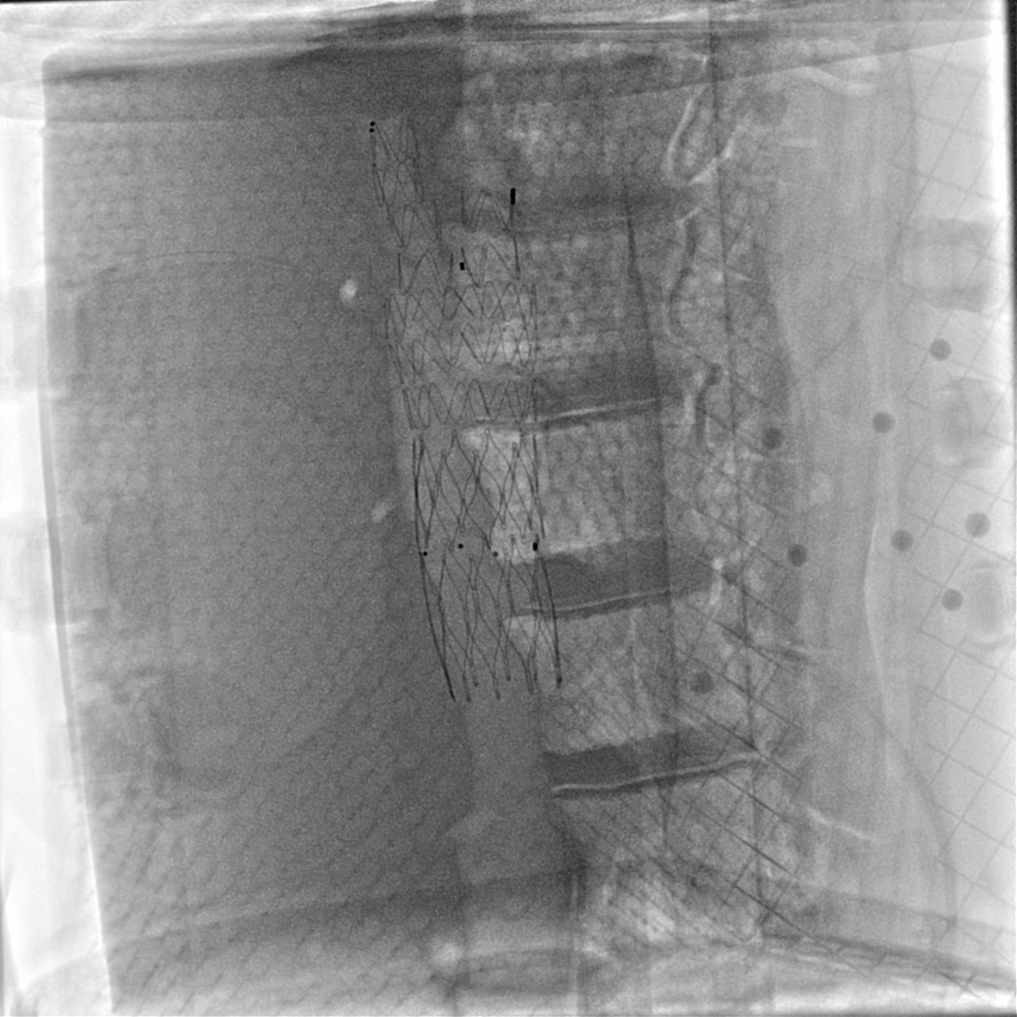}&
\includegraphics[height=.19\textwidth]{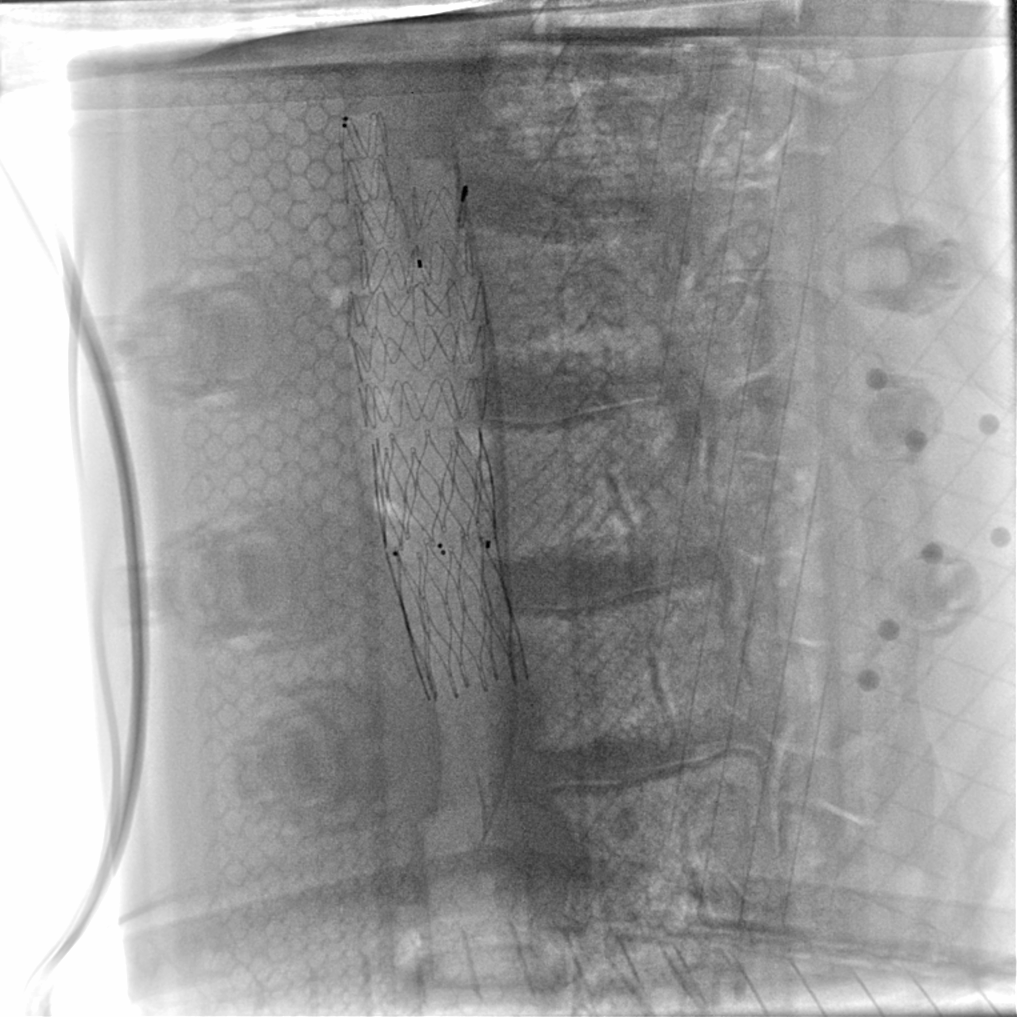}&
\includegraphics[height=.19\textwidth]{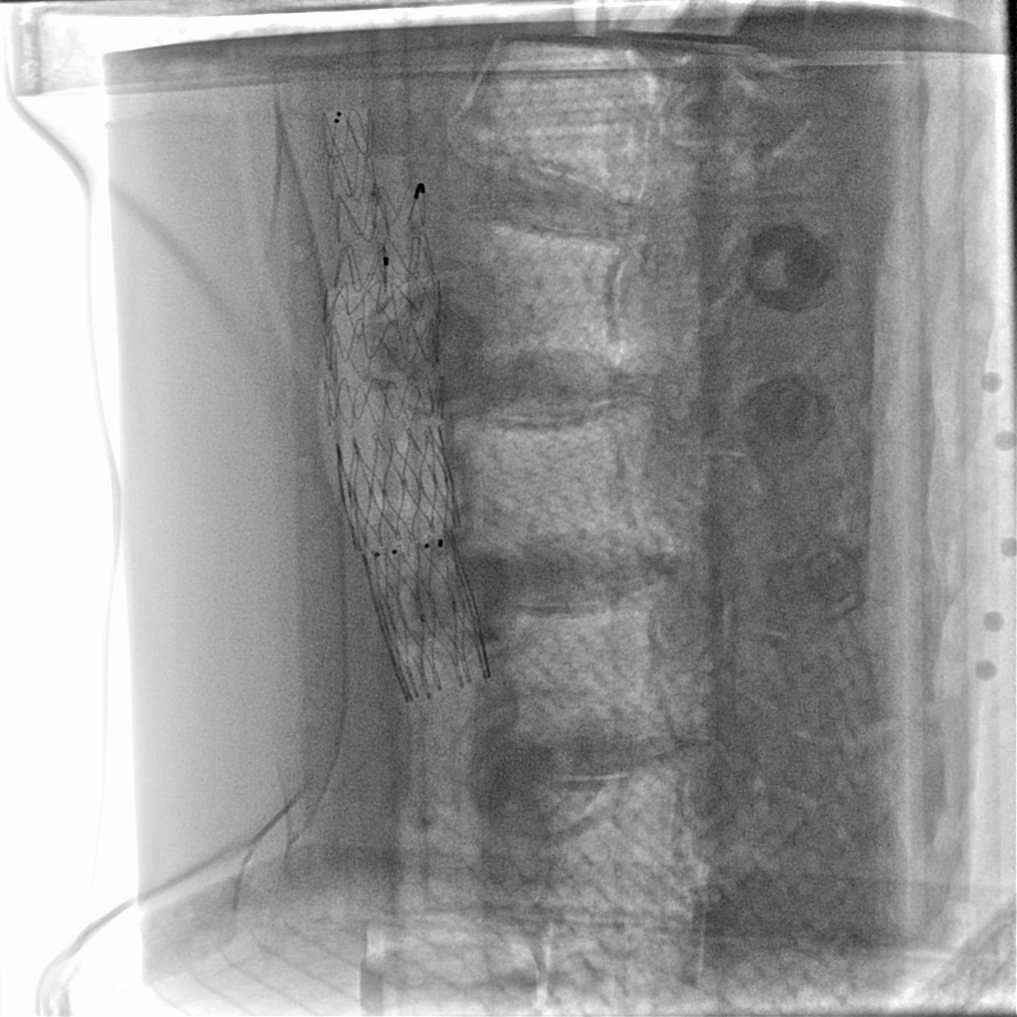}&
\includegraphics[height=.19\textwidth]{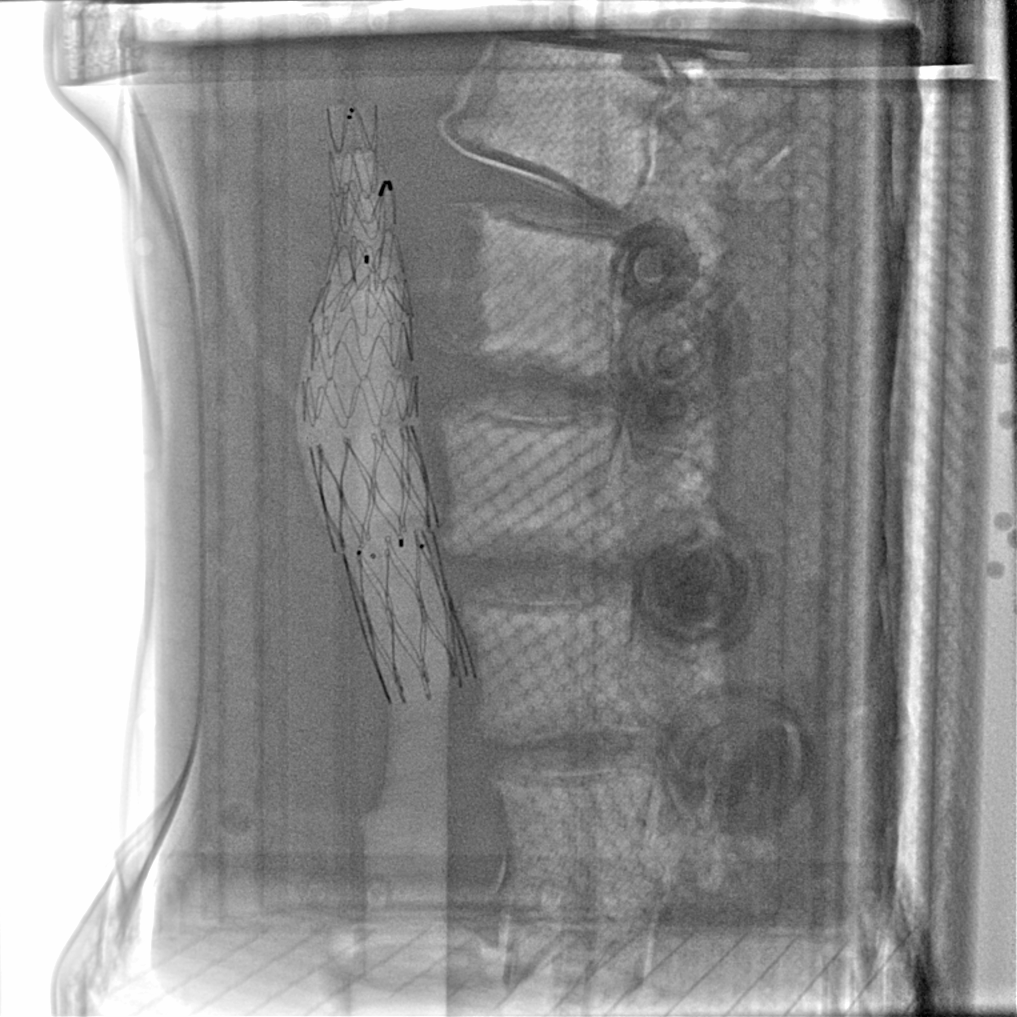}
\\

\includegraphics[width=.19\textwidth]{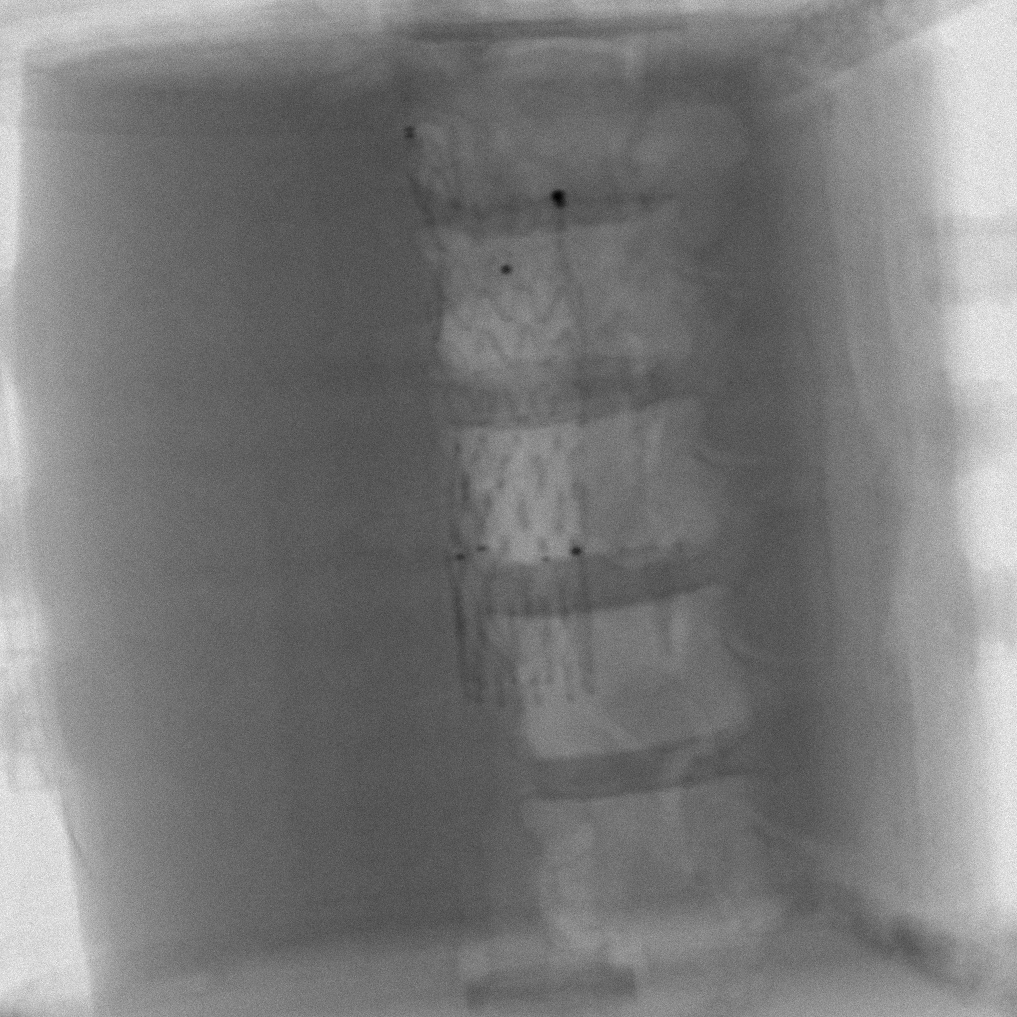}&
\includegraphics[width=.19\textwidth]{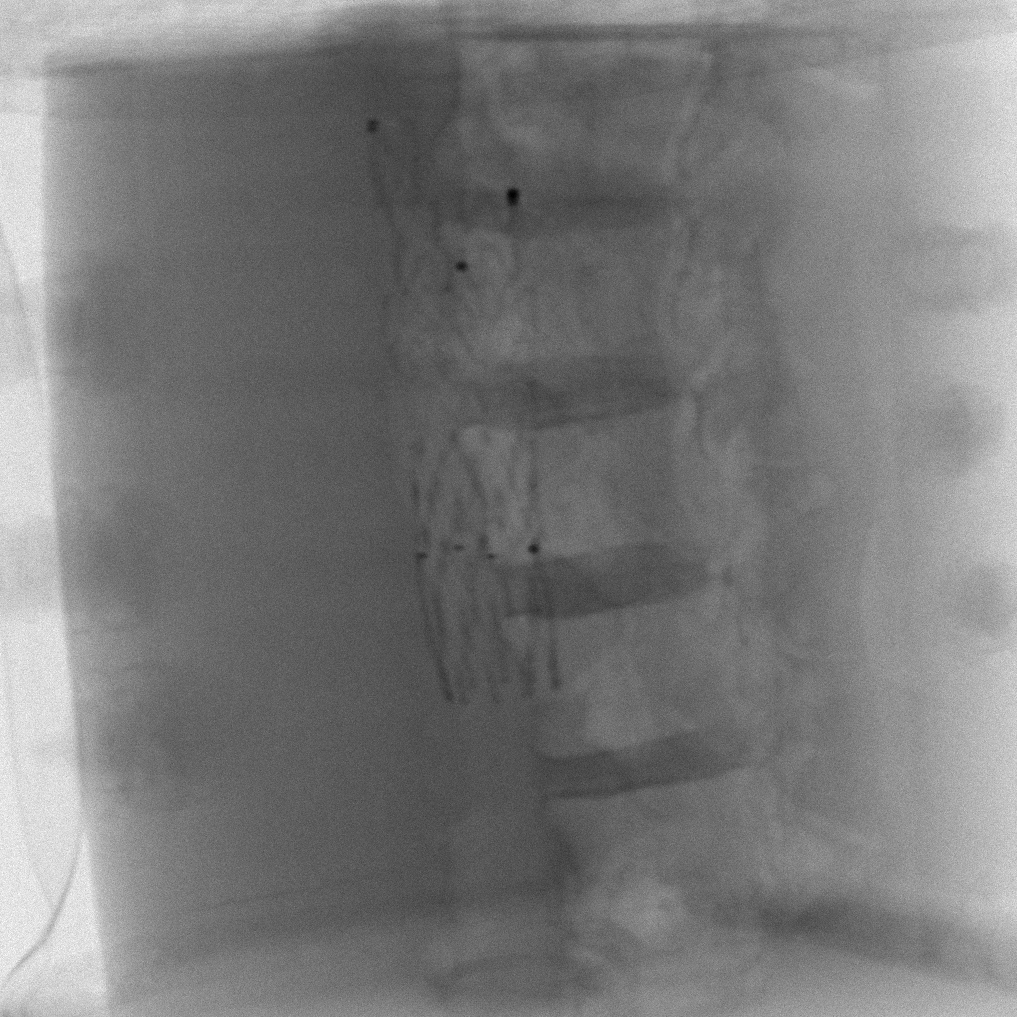}&
\includegraphics[height=.19\textwidth]{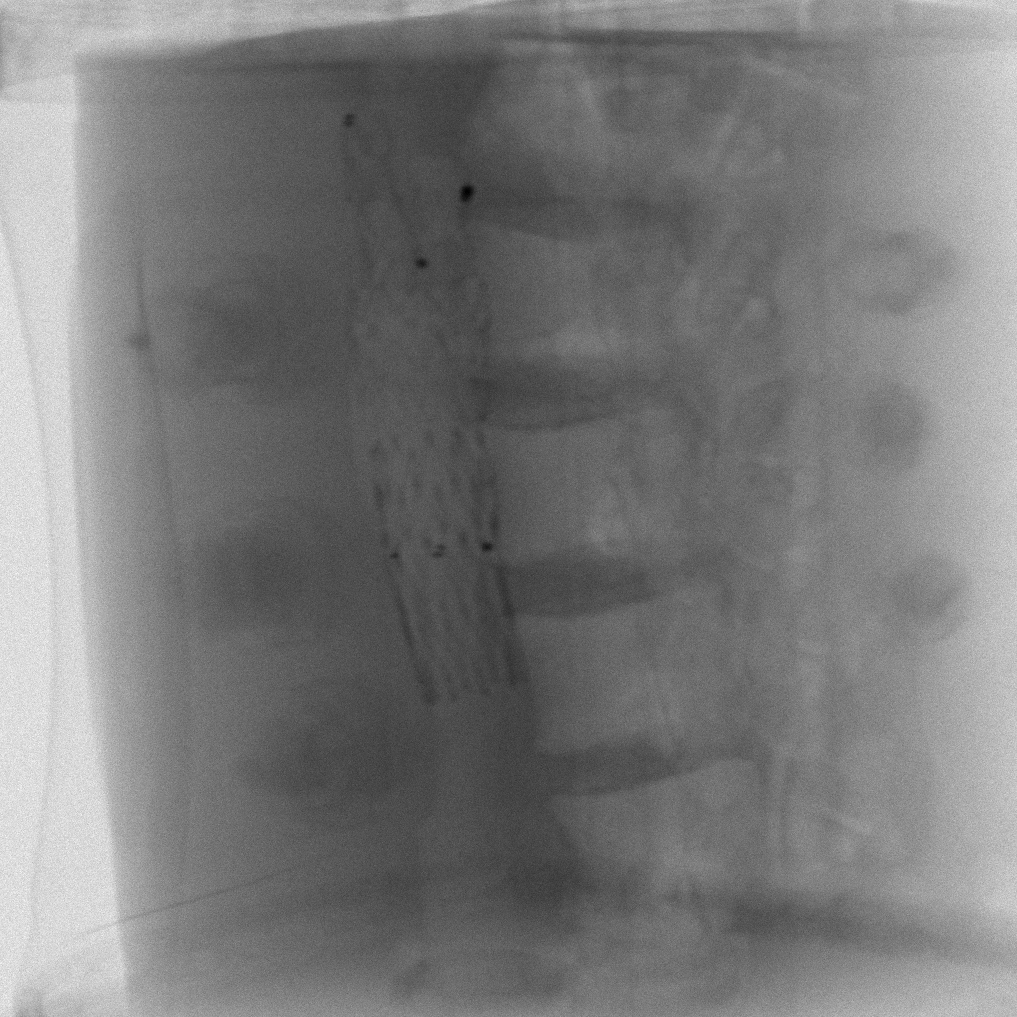}&
\includegraphics[height=.19\textwidth]{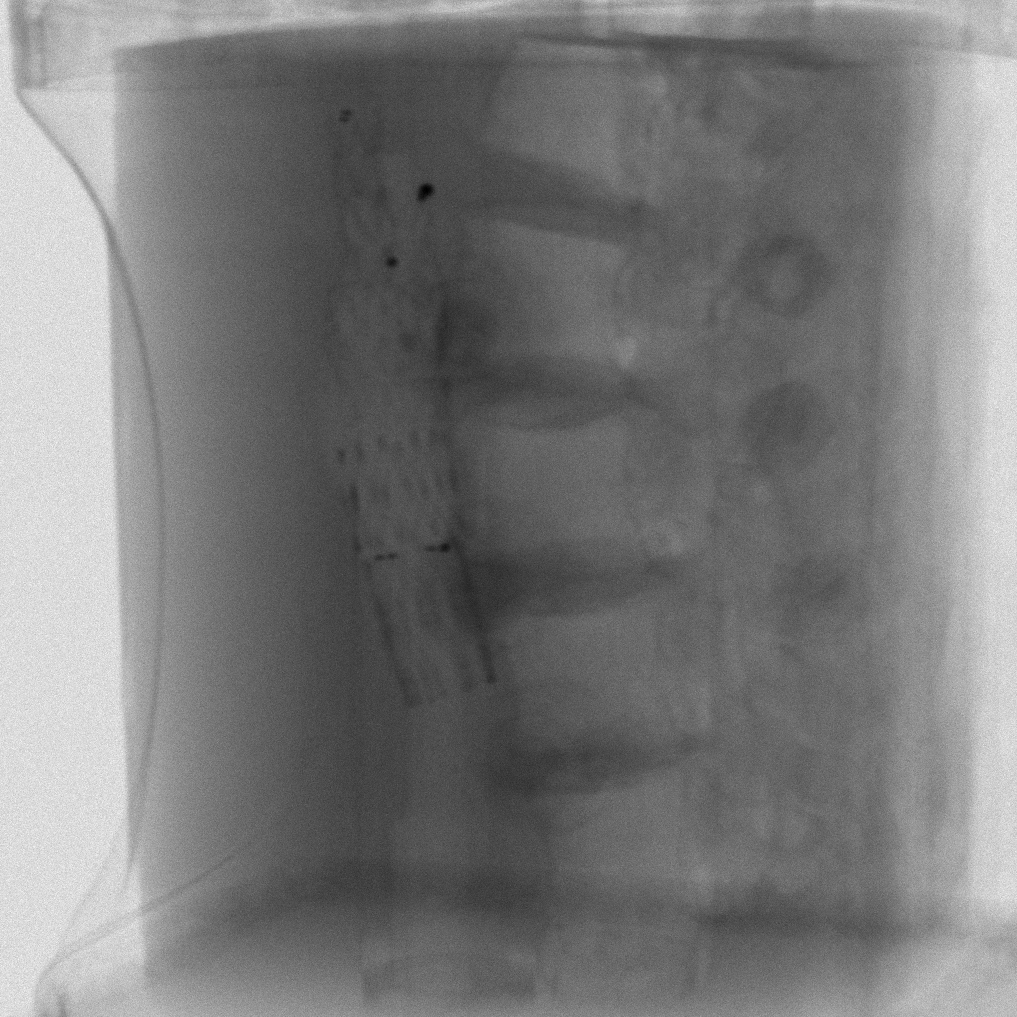}&
\includegraphics[height=.19\textwidth]{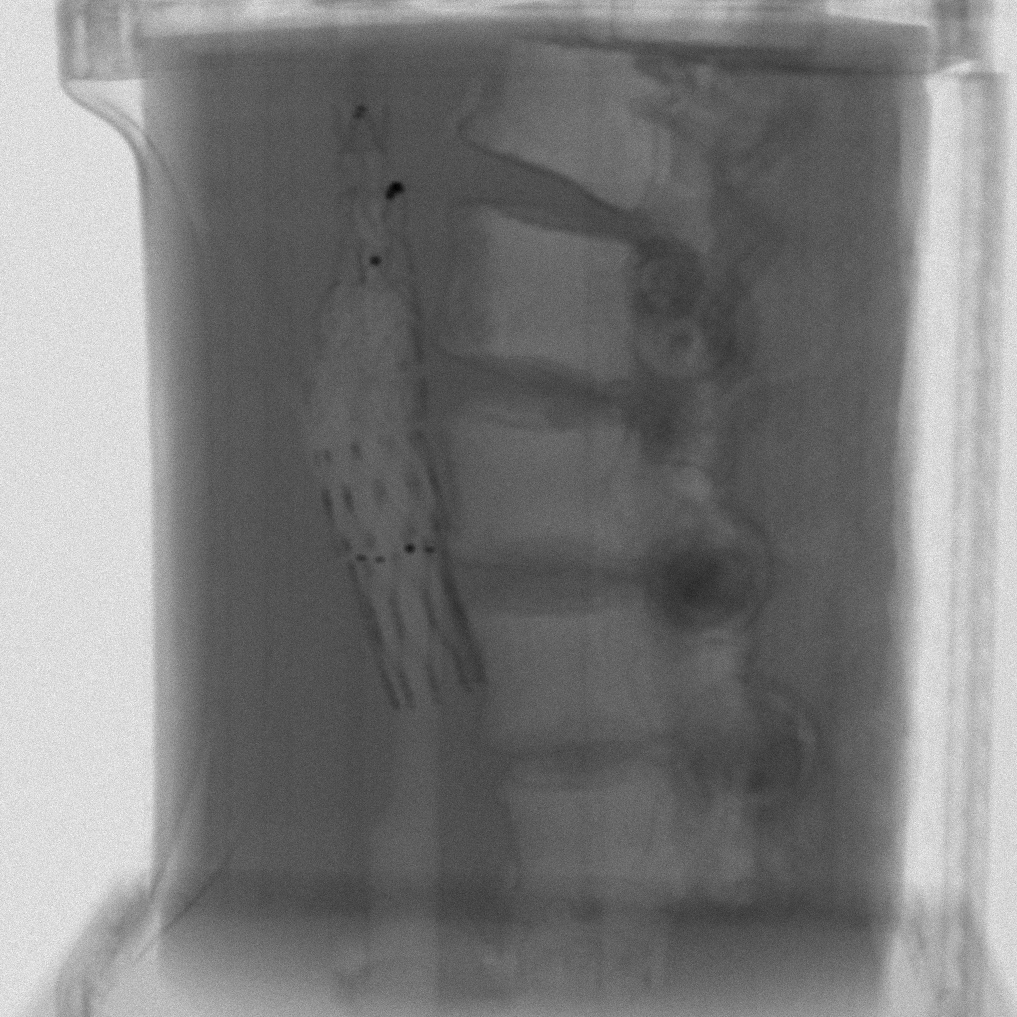}
\\

{LAO~\SI{10}{\degree}} & {LAO~\SI{30}{\degree}} & {LAO~\SI{50}{\degree}} & {LAO~\SI{70}{\degree}} & {LAO~\SI{90}{\degree}}
\\
%\multicolumn{3}{c}{C} & \multicolumn{3}{c}{D}\\
%\hline
\end{tabular}}
\caption{Results of C-arm virtual fluoroscopy on the \enquote{endoleak} phantom. (top row: real fluoroscopic images; bottom row: virtual fluoroscopic images)}
\label{fig_virtual_fluoroscopy}
\end{figure}

\begin{figure*}[htbp]
\centering
\begin{tabular}{c}
\includegraphics[width=.90\textwidth]{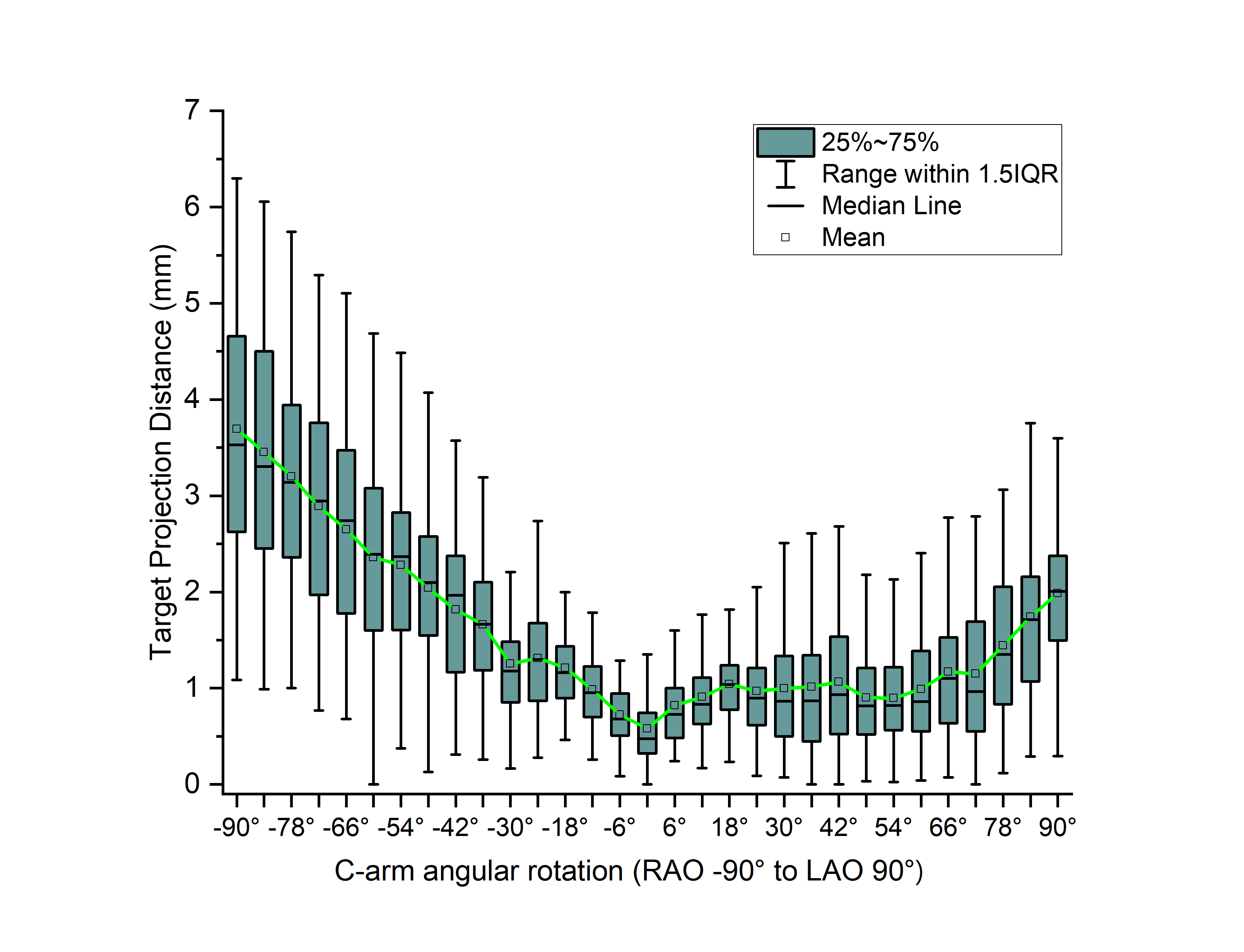}
\\
%{(A)}\\
\end{tabular}
\caption{Target mean projection distance error of virtual fluoroscopic images from RAO to LAO (\SI{-90}{\degree} to~\SI{90}{\degree}).}
\label{fig_target_projection_distance}
\end{figure*}

%\vspace{-0.5 cm}
We acquired 90 fluoroscopic images ranging from C-arm RAO to LAO (\SI{-90}{\degree} to \SI{90}{\degree}) in \SI{2}{\degree} increments to evaluate virtual fluoroscopy with joint encoder readings obtained from the Canon Alphenix C-arm (Canon Inc., Tokyo, Japan). In Fig.~\ref{fig_virtual_fluoroscopy}, the simulated fluoroscopic images presented a lower resolution than real fluoroscopic images, but the anatomy structure information and imaging perspective can be successfully simulated. Additionally, Fig.~\ref{fig_target_projection_distance} shows the mPDs between real and simulated fluoroscopic images, ranging from RAO to LAO (\SI{-90}{\degree} to ~\SI{90}{\degree}). An mPD of approximately~\SI{1.0}{\milli\metre} was achieved within \SI{-30}{\degree} to \SI{30}{\degree} C-arm range. The mPD increased when the C-arm moved further from the anterior-posterior (AP) view, likely due to the C-arm flex from gravity. Notably, high mPD values in virtual fluoroscopy (\eg, mPD $\approx$~\SI{3.5}{\milli\metre} at RAO~\SI{90}{\degree}) may affect precise C-arm repositioning, though this does not impact interventional procedure performance, which primarily depends on the accuracy of fluoro-CT registration and instrument tracking. 
% \subsection{Virtual Fluoroscopy}

%\begin{figure}[h]
%\centering
%\begin{tabular}{@{} c @{} c}
%\includegraphics[width=.5\textwidth]{figs/Target_projection_distance_vf.png}
%&{}
%\\
%{(A)}\\
%\end{tabular}
%\caption{Target mean projection distance error of virtual fluoroscopic images from RAO to LAO (\SI{-90}{\degree} to~\SI{90}{\degree})}
%\label{fig_target_projection_distance}
%\end{figure}

\begin{table*}[htbp]
\caption{Runtime performance of the virtual fluoroscopy workflow, tested on a system with an intel i7-9700K CPU and NVIDIA RTX 2080Ti GPU.}
\centering

\begin{tabular}{c|c}
        \hline
        {\makecell{Virtual fluoroscopy \\ workflow}} & {Runtime}\\
        \hline
        {\makecell{YOLO landmark \\ detection}} & {$\sim$38 fps}\\
        \hline
        {\makecell{Fluoro-CT \\ registration}} & {$\sim$1.97 s}\\
        \hline
        {DRR generation} & {\qty{0.15}{\second}--\qty{0.20}{\second}}\\
        \hline
        {Instrument tracking} & {$\sim$\qty{0.03}{\second} (30 fps)}\\
        \hline
        \end{tabular}
\label{tab_results_runtime}
\end{table*}

\subsection{Phantom Experiments}
\begin{figure}[h]
\centering
\begin{tabular}{@{} c @{} c @{}  c @{}  c @{}  c @{} c @{}}

%\hline
\includegraphics[trim={0cm 0cm 0cm 0cm}, clip, height=.16\textwidth]{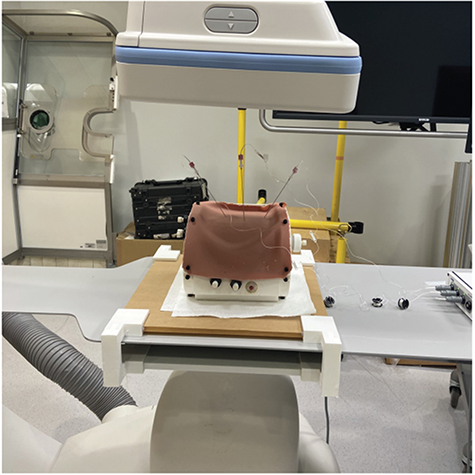}&
\includegraphics[trim={0cm 0cm 0cm 0cm}, clip,height=.16\textwidth]{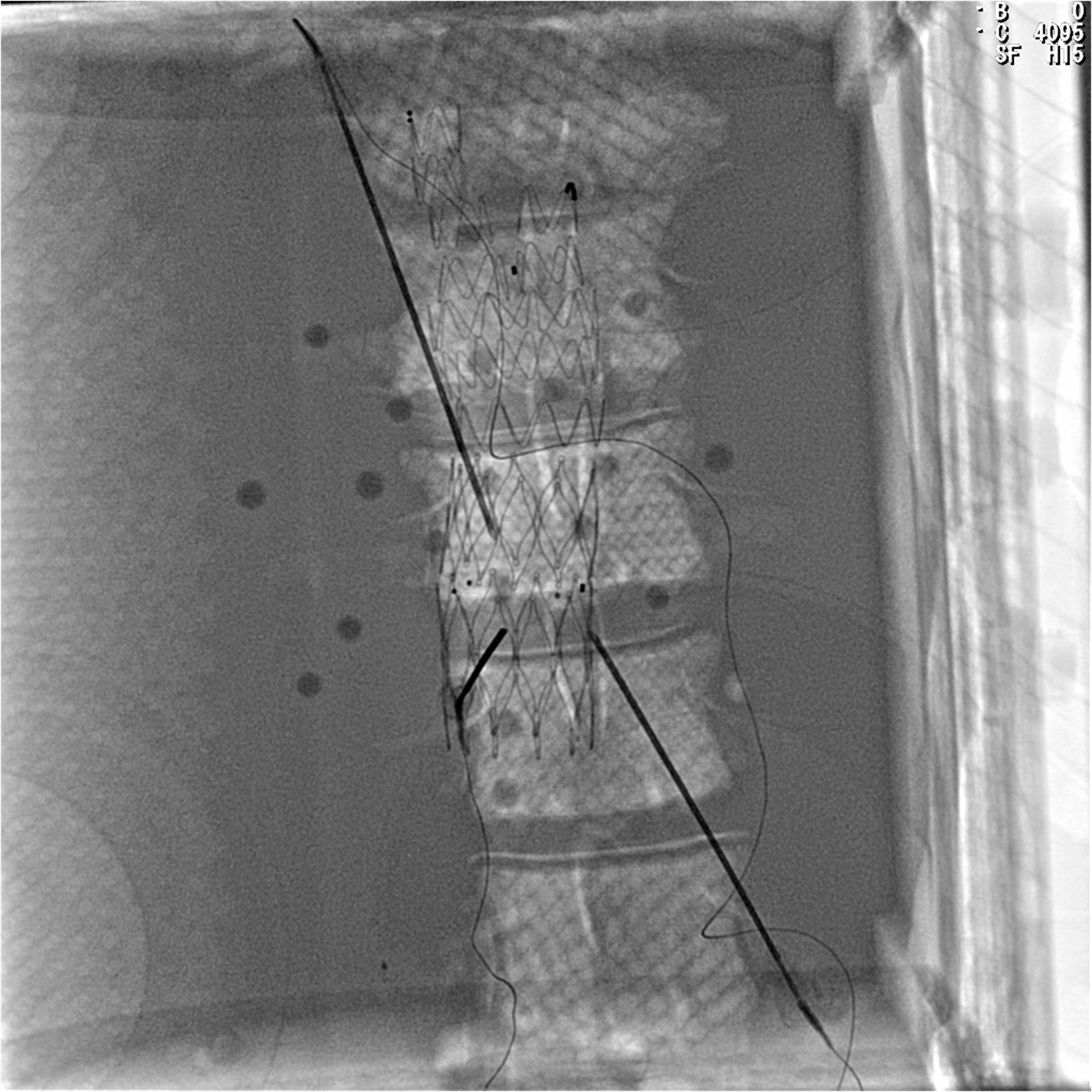}&
\includegraphics[trim={2cm 1.5cm 1.3cm 1.5cm}, clip, angle=270,origin=c,height=.16\textwidth]{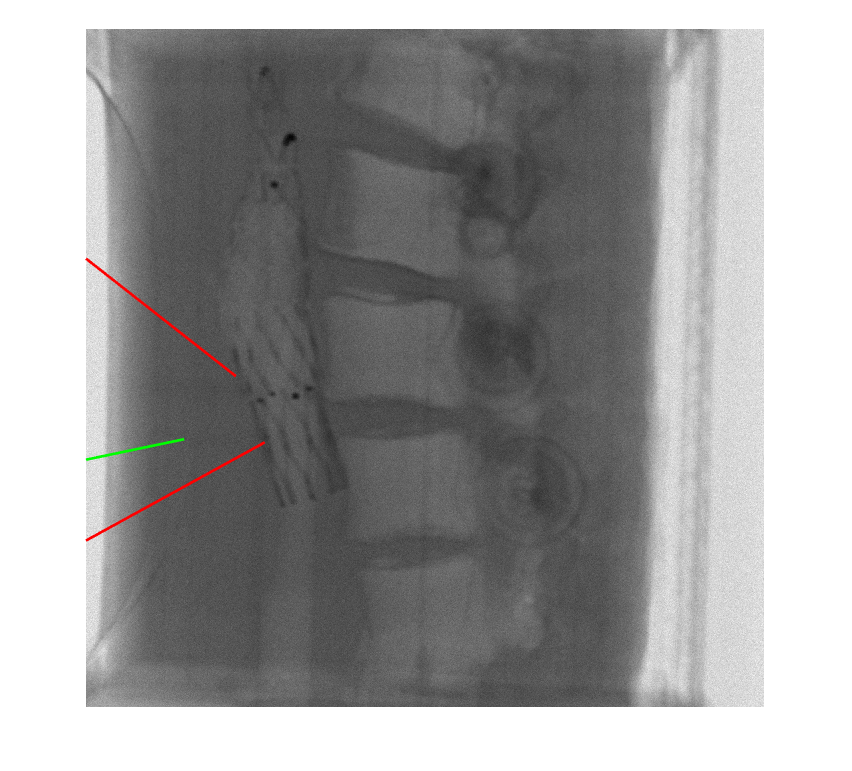}
&
\includegraphics[trim={2cm 1.5cm 1.3cm 1.5cm}, clip,angle=270,origin=c,height=.16\textwidth]{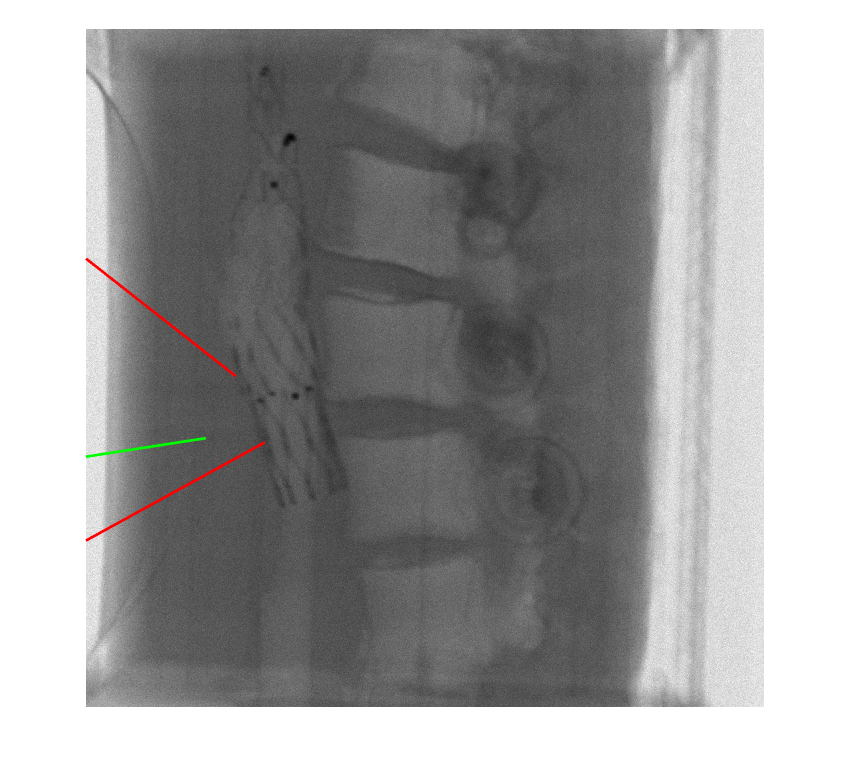}&
\includegraphics[trim={2cm 1.5cm 1.3cm 1.5cm}, clip,angle=270,origin=c,height=.16\textwidth]{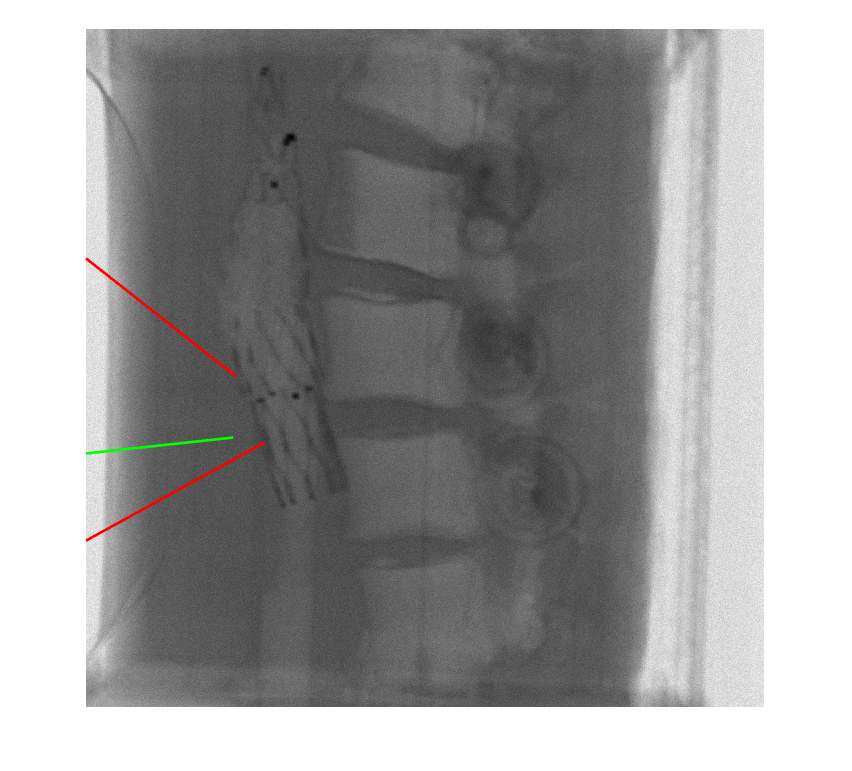}&
\includegraphics[trim={2cm 1.5cm 1.3cm 1.5cm}, clip,angle=270,origin=c,height=.16\textwidth]{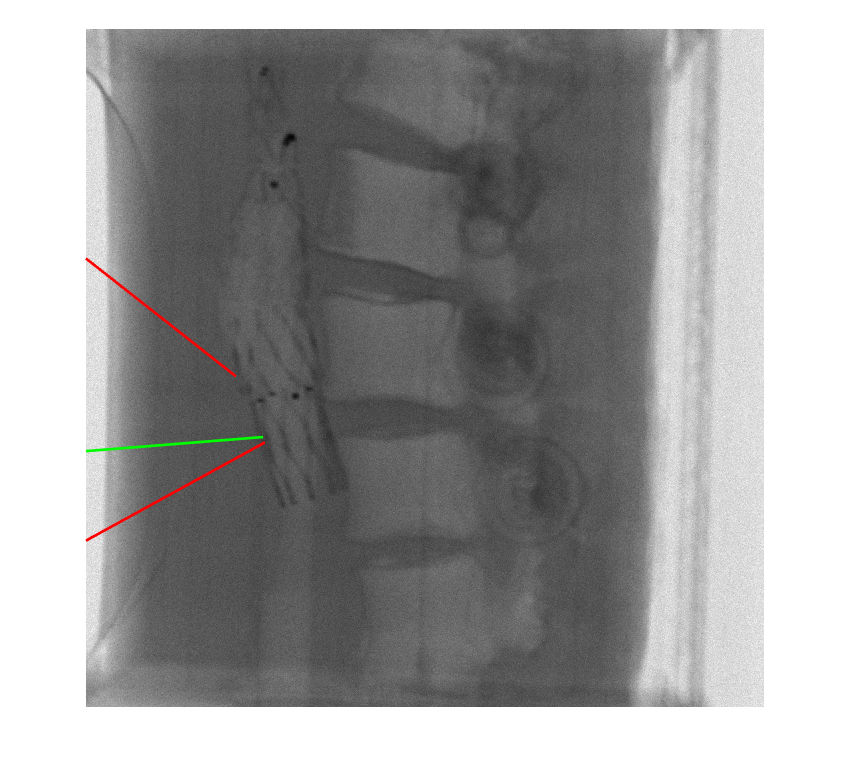}
\\
{Setup} &{AP} & {v-lateral-1} & {v-lateral-2} & {v-lateral-3} & {v-lateral-4}

\\
\includegraphics[trim={0cm 0cm 0cm 0cm}, clip,height=.16\textwidth]{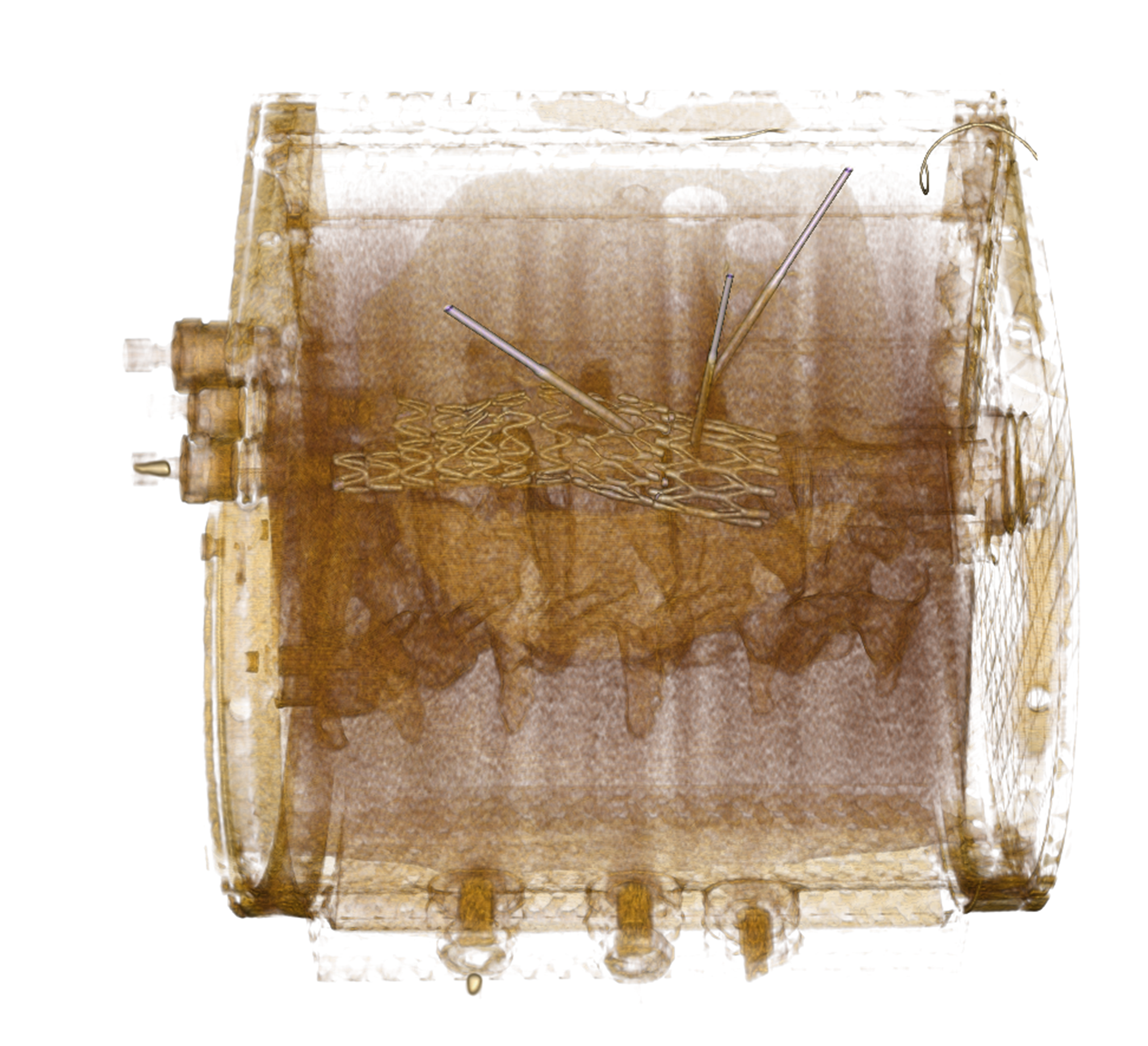}&
\includegraphics[trim={0cm 0cm 0cm 0cm}, clip,height=.16\textwidth]{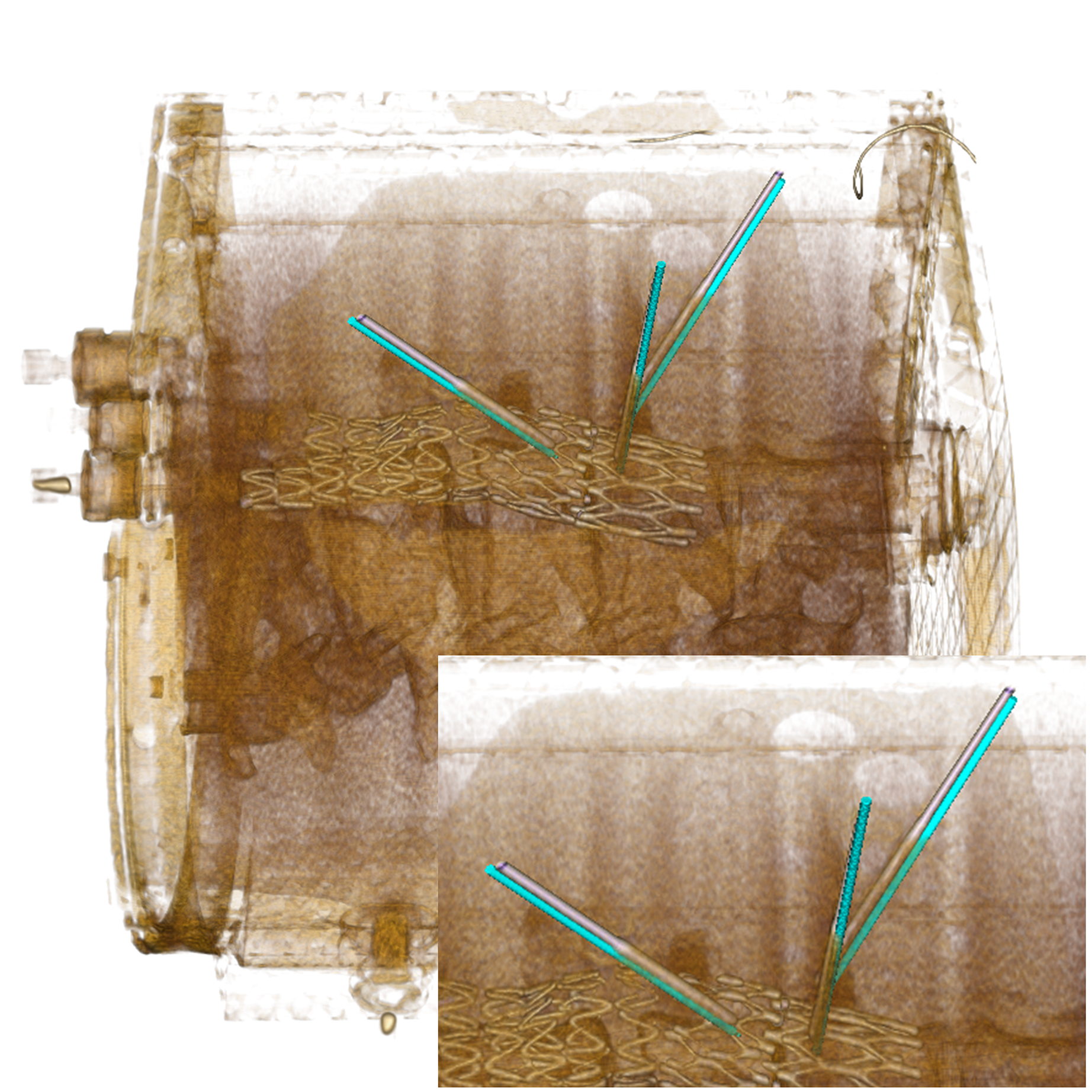}&
\includegraphics[trim={2cm 1.5cm 1.3cm 1.5cm}, clip,angle=270,origin=c,height=.16\textwidth]{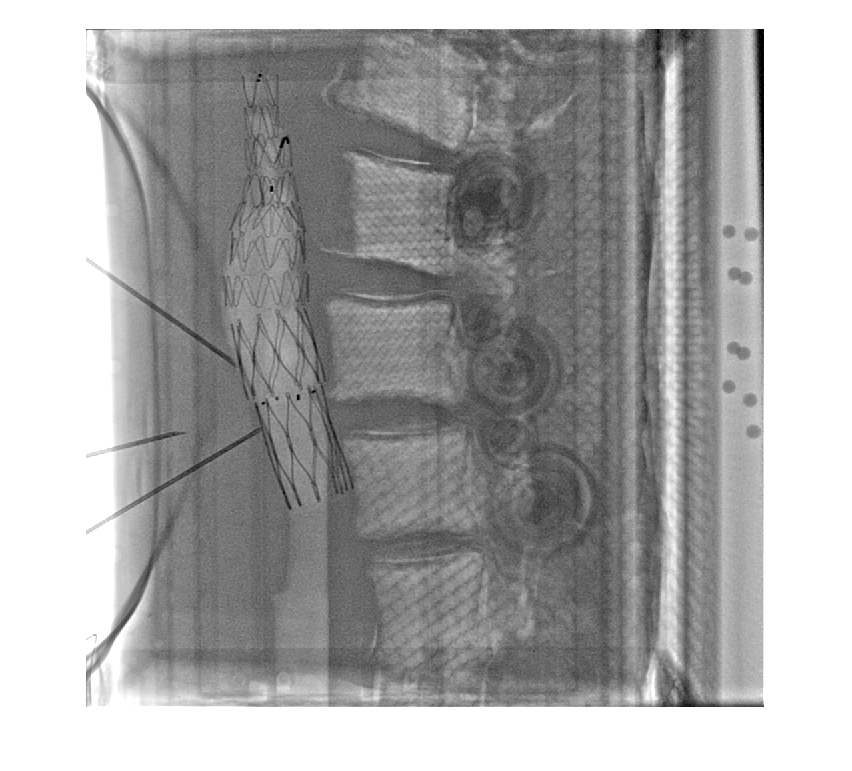}&
\includegraphics[trim={2cm 1.5cm 1.3cm 1.5cm}, clip,angle=270,origin=c,height=.16\textwidth]{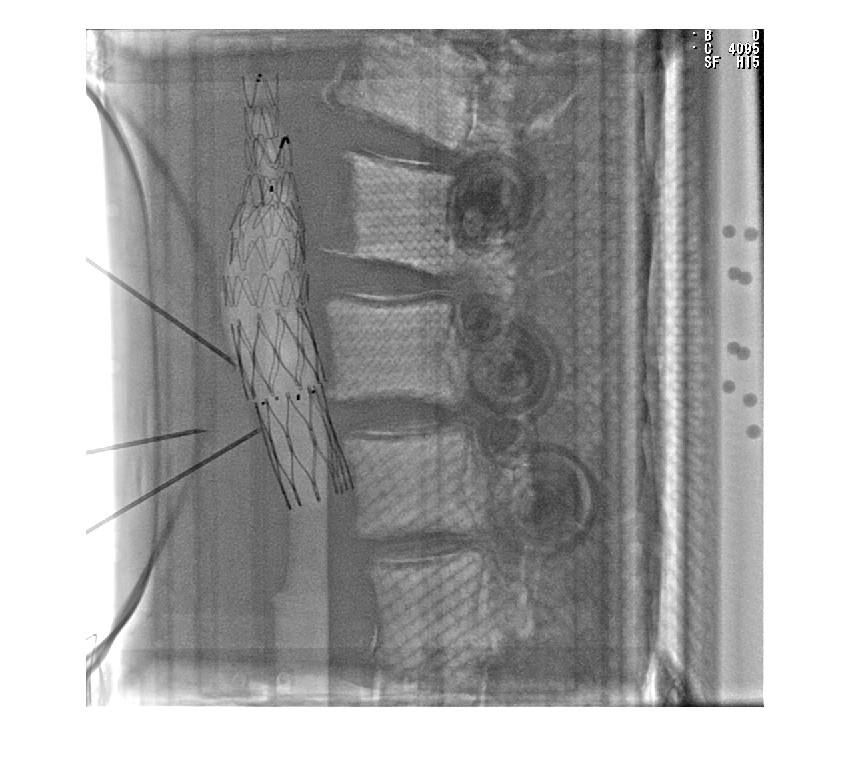}&
\includegraphics[trim={2cm 1.5cm 1.3cm 1.5cm}, clip,angle=270,origin=c,height=.16\textwidth]{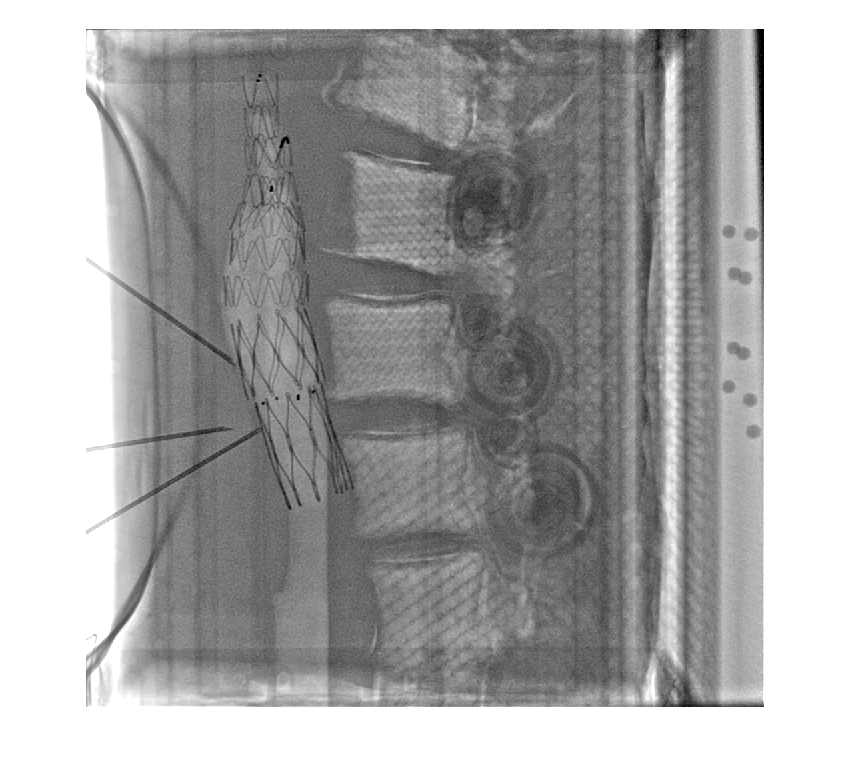}&
\includegraphics[trim={2cm 1.5cm 1.3cm 1.5cm}, clip,angle=270,origin=c,height=.16\textwidth]{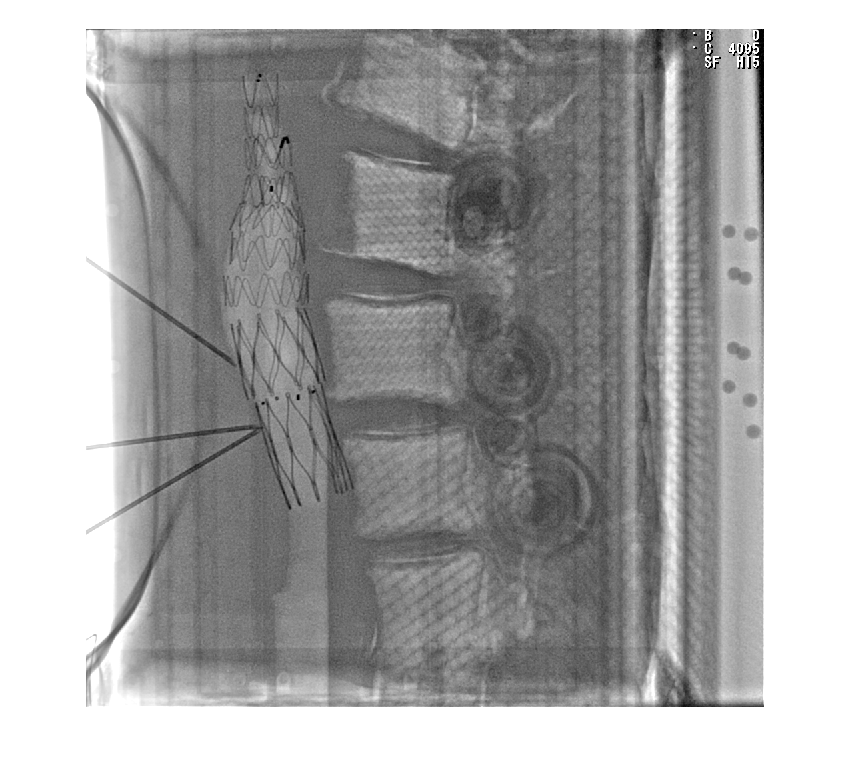}
\\
{3D-1}&{3D-2} & {lateral-1} & {lateral-2} & {lateral-3} & {lateral-4}
\\

%\multicolumn{3}{c}{C} & \multicolumn{3}{c}{D}\\
%\hline
\end{tabular}
\caption{Phantom insertion experiments using virtual fluoroscopy and 3D navigation. In simulated images, the advancing and inserted needles are in green and red, respectively. In 3D view, the virtual and real needles are in cyan and silver, respectively. (v-lateral-\#: virtual lateral view \#, lateral-\#: real lateral view \# )}
\label{fig_virtual_fluoroscopy_phantom_targeting}
\end{figure}

A phantom insertion experiment was conducted (see Fig.~\ref{fig_virtual_fluoroscopy_phantom_targeting}\textcolor{black}{-Setup}) to evaluate the efficacy of virtual fluoroscopy in providing multiplanar fluoroscopic views for facilitating needle insertion. The real and \textcolor{black}{virtual} lateral views (see Fig.~\ref{fig_virtual_fluoroscopy_phantom_targeting}\textcolor{black}{-lateral-\#}) showed visually imperceptible differences in representing the phantom structure and needle positions. To assess needle insertion accuracy, we acquired a phantom CT image after the needles were inserted (see Fig.~\ref{fig_virtual_fluoroscopy_phantom_targeting}\textcolor{black}{-3D-1} in silver), the magnetically tracked needles were overlaid onto the registered CT image (see Fig.~\ref{fig_virtual_fluoroscopy_phantom_targeting}\textcolor{black}{-3D-2} in cyan). Needle tip and angle errors were calculated by comparing the tracked needles with their segmented CT counterparts.
The needle tip errors were \SI{3.46}{\milli\metre}, \SI{2.87}{\milli\metre} and \SI{3.95}{\milli\metre}, while orientation errors were \SI{1.76}{\degree}, \SI{1.02}{\degree} and \SI{0.13}{\degree}. \textcolor{black}{Tab.~\ref{tab_results_runtime} summarizes the workflow’s runtime performance from the phantom insertion experiment. The experimental setup used a phantom CT image with dimensions of \numproduct{365 x 280 x 791} pixels and spacings of \qtyproduct{0.71 x 0.71 x 0.30}{\milli\metre}, and fluoroscopic images with dimensions of \numproduct{1017 x 1017} pixels. Fluoro-CT registration required $\sim$\qty{1.97}{\second} per image pair. DRR generation achieves runtimes of \qty{0.15}{\second}-\qty{0.20}{\second} per image, which is deemed clinically acceptable for fluoroscopy guidance.}

%\begin{table}[htbp]
%\caption{Runtime performance of the virtual fluoroscopy workflow, tested on a system with an intel i7-9700K CPU and NVIDIA RTX 2080Ti GPU.}
%\centering
%\setlength{\tabcolsep}{8pt}
%\begin{tabular}{cc}
%\hline
%{Virtual fluoroscopy workflow} & {Runtime/frame rate}\\
%\hline
%{YOLO landmark detection} & {$\sim$30 fps}\\
%{Fluoro-CT registration} & {$\sim$1.97 s}\\
%{DRR generation} & {\qtyrange{0.15}{0.20}{\second}}\\
%{Instrument tracking} & {$\sim$\qty{0.03}{\second} (30 fps)}\\
%\hline
%\end{tabular}
%\label{tab_results_runtime}
%\end{table}

%\vspace{-1.0 cm}

\section{Discussion}\label{discussion}

\textcolor{black}{A virtual fluoroscopy workflow, combined with the integration of magnetic tracking, was introduced for} improving depth perception and reducing the need for repeated fluoroscopic image acquisition. 1) \textcolor{black}{Conventionally}, displaying multiple fluoroscopic views, typically AP and lateral views, helps the radiologist appreciate 3D anatomical structures. Our virtual fluoroscopy allows multiplanar views to be visualized simultaneously without frequent C-arm repositioning, thereby reducing the radiologist or technician's workload during procedures. \textcolor{black}{Combining the consistent tracking performance and the generalized fluoro-CT registration approach, the guidance accuracy ($\sim$\SI{3.0}{\milli\metre}) demonstrated on the phantom is, in theory, representative of other clinical applications.} Given that the DRR approach theoretically does not introduce additional projection errors in simulated images, this $\sim$\SI{3.0}{\milli\meter} error also reflects the accuracy of virtual fluoroscopy. \textcolor{black}{Clinically, the observed insertion performance is suitable for a wide range of procedures, such as biopsies and vascular access interventions such as endovascular aneurysm repair~\cite{manstad2011three} and endoleak repair~\cite{hefel1995internal}. However, additional experiments are required to validate the feasibility for procedures that demand higher accuracy, such as pedicle screw placement~\cite{aoude2015methods} and vertebroplasty~\cite{noguchi2022accuracy}.} 2) Fig.~\ref{fig_virtual_fluoroscopy} demonstrates that our virtual fluoroscopy is capable of accurately capturing a wide range from RAO to LAO (\SI{-90}{\degree} to \SI{90}{\degree}), showing potential in improving users' understanding of fluoroscopy imaging principles, thereby reducing repeated acquisitions. Although the mPD of virtual fluoroscopy increased as the C-arm moves away from the AP view due to gantry flexing (see Fig.~\ref{fig_virtual_fluoroscopy}), phantom targeting experiments demonstrated that the simulated lateral views did not have a clinically significant impact on interventions. Consequently, the number of fluoroscopic acquisitions could be reduced during both C-arm repositioning and instrument guidance. Future work will focus on quantitatively evaluating the reduction in radiation exposure facilitated by virtual fluoroscopy compared to conventional workflows. 

In fluoroscopy-guided procedures, there are two types of C-arm imaging systems: manually driven and mechanically driven. Mechanically driven systems, often used in complex vascular and cardiac procedures, feature motorized and tracked movements. \textcolor{black}{Our approach is compatible with both manually and mechanically driven C-arms for providing multiplanar views and enhancing 3D visualization. Additionally, we demonstrated the efficacy of generating virtual fluoroscopic images corresponding to the actual C-arm pose, validated on a mechanically driven C-arm. Extending this capability to manually driven C-arms is in principle possible~\cite{grzeda2010c} but requires further exploration.}

\textcolor{black}{The added complexity of the proposed approach must be justified by the clinical benefits. 1) The use of fiducials.} To eliminate \textcolor{black}{image} artifacts necessitated by the aluminum fiducials and to a lesser extent those created by the radiolucent FG prototype, we developed a Deep Adversarial Decomposition approach, as detailed in \cite{xia2023x}, which effectively removes artifacts without affecting guidance. \textcolor{black}{ 2) Impact on clinical workflows. The FG mounting frame's two-layer design enables seamless installation on the surgical bed during preparation, without disrupting conventional setup steps. Additionally, a ``server-client'' framework has been introduced for streaming, processing, and visualizing multimodal data~\cite{xing2024towards}. Our current development addresses setup complexity and fiducial management through targeted solutions. Future work will evaluate the system in real clinical settings to further optimize its usability.}

\textcolor{black}{Our virtual fluoroscopy approach demonstrated feasibility for scenarios where imaged objects can be reasonably assumed as globally rigid bodies. The generalization capability of our fluoro-CT registration was evaluated in~\cite{xing2024towards}. However, in clinical cases (\eg, bone fracture reduction), relative motion between imaged objects may result in inaccuracies in virtual fluoroscopic images due to reliance on non-updated preoperative CT images. Grupp~\etal~\cite{grupp2019pose} proposed an X-ray-CT registration approach in periacetabular osteotomies, employing a strategy to sequentially align the pelvis, femur, and acetabular fragment. Similarly, Gao~\etal~\cite{gao2020fiducial} used this registration strategy to recover the relative motion between the pelvis and femur in femoroplasty. Therefore, this sequential registration strategy holds potential for determining the relative pose of imaged objects, enabling the generation of accurate virtual fluoroscopic images}

\section{Conclusion}\label{conclusion}
Based on \textcolor{black}{the integration of} magnetic tracking into fluoroscopy-guided workflows, we proposed an efficient workflow to achieve virtual fluoroscopy, including robust fluoro-CT registration, automatic 2D-3D landmark correspondence, and a general C-arm pose modelling. The broad capture range and accuracy of our virtual fluoroscopy approach showed promise in improving the users' understanding of X-ray imaging principles, thereby reducing the number of fluoroscopy acquisitions during procedures. Phantom targeting experiments demonstrated the efficacy of simulated multiplanar views in providing depth information for guidance.

%%===========================================================================================%%
%% If you are submitting to one of the Nature Portfolio journals, using the eJP submission   %%
%% system, please include the references within the manuscript file itself. You may do this  %%
%% by copying the reference list from your .bbl file, paste it into the main manuscript .tex %%
%% file, and delete the associated \verb+\bibliography+ commands.                            %%
%%===========================================================================================%%
\section*{Declarations}

\textbf{Funding:} This work was supported in part by the Canadian Institutes of Health Research (CIHR) Foundation Grant 201409, in part by the Natural Sciences and Engineering Research Council (NSERC) of Canada under Grant 2024-06674, in part by the Canadian Foundation for Innovation under Grant 36199, in part by INOVAIT, in part by Canon Medical Systems.

\noindent\textbf{Competing Interests:} The authors declare that they have no conflict of interest.

\noindent\textbf{Ethics:} No ethical approval is needed for this study.

\noindent\textbf{Informed Consent:} No informed consent is required.

\bibliography{sn-bibliography}% common bib file
%% if required, the content of .bbl file can be included here once bbl is generated
%%\input sn-article.bbl

\end{document}